# A Hybrid Simulation Code for Hall Thrusters and its Sensitivity to Numerical Parameters

Xingdong Che[a], Hong Li[a,b,*], Xin Guo[c], Zhaoyu Wang[d,e], Xifeng Cao[f], Daren Yu[a,b]

[a] Lab of Plasma Propulsion, Harbin Institute of Technology, Harbin 150001, People's Republic of China.

[b] Key Laboratory of Advanced Aerospace Propulsion Technology, Ministry of Industry and Information Technology, Harbin 150001, People's Republic of China.

[c] Aviation Key Laboratory of Science and Technology on Aerodynamics of High Speed and High Reynolds Number, AVIC Aerodynamics Research Institute, Shenyang 110034, People's Republic of China.

[d] China Jiuyuan Hi-Tech Equipment Co., Ltd., Beijing 100094, People's Republic of China.

[e] CAEP Software Center for High Performance Numerical Simulation, Beijing 100088, People's Republic of China.

[f] School of Mechanical and Power Engineering, Harbin University of Science and Technology, Harbin 150001, People's Republic of China.

*Corresponding author.
E-mail address: lihong@hit.edu.cn (Hong Li)

**Abstract**

Hybrid methods offer an attractive balance between computational efficiency and physical accuracy, playing an important guiding role in the design and optimization of Hall thrusters. In this work, a hybrid simulation code, named HYSCH, is developed, utilizing a 1D-MFAM as the mesh framework for the electron submodel. HYSCH enables decoupling of the two submodel meshes, providing a higher degree of flexibility in spatial resolution settings. The sensitivity of the simulated results to numerical parameters is analyzed to validate the model's **numerical robustness and physical consistency.** The analysis shows that the temporal resolution of the heavy-species submodel and the spatial resolution of the electron submodel exert a more pronounced influence on the agreement with measurements than the heavy-species macroparticle weight and spatial resolution, confirming that the decoupled formulation enables more efficient resource allocation than coupled discretizations.

## 1 Introduction

Hall thrusters have become a widely adopted electric-propulsion technology for a broad range of space missions because of their high specific impulse and favorable propulsive efficiency. The increasing diversity of mission requirements has stimulated the development of numerous Hall-thruster configurations, while experimental development and testing have become increasingly costly[1]. Consequently, there is a growing need for high-fidelity numerical models that can resolve the coupled ionization and acceleration processes and support thruster design and optimization[2,3]. Existing numerical approaches can be broadly classified as particle-in-cell (PIC), fluid, and hybrid methods[2,3]. Among these approaches, hybrid methods provide an attractive compromise between computational efficiency and physical fidelity by treating heavy species kinetically as discrete particles while describing electrons as a continuum fluid. Compared with fully kinetic PIC methods, hybrid methods substantially reduce the computational cost by eliminating the need to resolve electron kinetic scales. Compared with purely fluid methods, they provide a more accurate representation of nonequilibrium ion velocity distributions[4].

A major challenge in the hybrid modeling of Hall thrusters arises from the strong magnetization of electrons, which results in highly anisotropic particle and energy transport. In some regions, the electron conductivity parallel to the magnetic field may exceed the cross-field conductivity by four to five orders of magnitude. Such strong anisotropy makes the numerical solution of the electron equations highly sensitive to the alignment between the computational mesh and the magnetic field. Pérez-Grande et al.[5] demonstrated that mesh misalignment with respect to the magnetic-field direction can introduce substantial numerical diffusion and significantly degrade simulation accuracy. These considerations motivate the use of a magnetic-field-aligned mesh (MFAM), together with separate treatments of electron transport parallel and perpendicular to the magnetic field.

The development of MFAM-based Hall-thruster models can be traced back to 1995. Komurasaki et al.[6] reported an early hybrid model based on a one-dimensional magnetic-field-aligned mesh (1D-MFAM). In their formulation, ion trajectories were advanced using Newton's equations of motion, while the relevant electron momentum- and energy-conservation equations were solved along magnetic field lines. Fife developed a Hall-thruster-specific 1D-MFAM framework by establishing relationships among the magnetic stream function, magnetic field lines, and plasma quantities and solving the resulting quasi-one-dimensional electron-fluid equations[7,8]. A coupled nonuniform structured mesh was constructed for the heavy-species submodel, leading to the development of HPHall, which subsequently became an influential reference for Hall-thruster hybrid models. During approximately the same period, Hagelaar et al.[9,10]and Koo et al.[11] developed related hybrid models that also employed a 1D-MFAM for the electron submodel, with electron collisions and energy losses represented primarily through empirical closures.

Parra et al.[12] subsequently developed HPHall-2 on the basis of HPHall by introducing more physics-based descriptions of electron particle and energy transport. They also implemented corrections to improve the representation of plasma–wall and sheath interactions, thereby enhancing the predictive capability of the model[12–17]. In addition, they incorporated a hot thermionic-emitter boundary condition and extended the framework into HallMA, a model developed specifically for two-stage Hall thrusters[18]. Further improvements were introduced by other researchers. For example, Hofer et al.[19] revised the thin-sheath wall model to account for the incomplete magnetization of electrons within the sheath, while Garrigues et al. [20] and Panelli et al.[21] reported closely related modeling developments.

Building on HPHall-2, Mikellides et al. incorporated magnetic equipotential lines into the field-aligned formulation to construct a two-dimensional magnetic-field-aligned mesh (2D-MFAM)[22,23]. By explicitly resolving electron thermal conduction parallel to the magnetic field, they developed the Hall2De electron-fluid model. Pérez-Grande et al.[5] subsequently demonstrated that the use of a 2D-MFAM can effectively suppress the numerical diffusion caused by mesh–field misalignment. The hybrid code HYPHEN was later developed using a 2D-MFAM and has since been applied to several Hall-thruster configurations, including the SPT-100, HT5K, HT20K, and CHT-200[24–28]. Following a similar strategy, Shashkov et al.[29] and Jung et al.[30] developed the Hybrid2D and MAHPS codes, respectively. Both codes employ triangular meshes for the heavy-species submodel while retaining a 2D-MFAM for the electron submodel.

Despite these advances, hybrid simulations based on a 1D-MFAM remain of considerable practical interest. A 2D-MFAM allows the electron model to resolve energy transport parallel to the magnetic field and departures from complete thermalization. Nevertheless, simulations often predict that the electron temperature remains nearly uniform along individual magnetic field lines, as observed in HYPHEN simulations of the HT5K Hall thruster[24–27]. This behavior suggests that simplified formulations based on near-thermalization along magnetic field lines may remain effective for many operating regimes. In conventional 1D-MFAM hybrid models, the heavy-species submodel is commonly discretized on a coupled nonuniform mesh. However, the relatively coarse spatial resolution frequently employed in the plume region can result in insufficient macroparticle sampling and, consequently, substantial physical and numerical errors in the deposited particle moments. To address this limitation, Koo et al.[11] employed a uniform structured mesh for the heavy-species submodel and decoupled its mesh topology from that of the electron submodel. Although their results were shown to be broadly insensitive to several numerical parameters, the effects of these parameters on numerical convergence were not quantitatively assessed.

The decoupling of heavy-species and electron meshes offers several important advantages from both physical and numerical perspectives. Physically, the optimal mesh alignment differs fundamentally between the two models: the electron transport is governed by highly anisotropic diffusion along magnetic field lines, whereas heavy-species transport is determined by the electric field and exhibits no preferential alignment with the magnetic field. A coupled mesh constrains both submodels to share the same spatial resolution and topology, which may either under-resolve ion dynamics in regions of strong electric fields or introduce excessive computational cost in regions where coarse electron resolution would suffice. Mesh decoupling allows each submodel to adopt its own optimal discretization, thereby improving the balance between numerical accuracy and computational efficiency. Furthermore, independent mesh generation simplifies the implementation of adaptive refinement strategies and facilitates sensitivity studies that isolate the numerical behavior of individual submodels. Despite these potential benefits, the robustness and physical consistency of decoupled-mesh hybrid models have not been systematically evaluated.

Motivated by these considerations, the present study develops a hybrid simulation code for Hall thrusters, referred to as HYSCH. In HYSCH, the heavy-species submodel is discretized on a uniform structured mesh, whereas the electron submodel is solved on a 1D-MFAM. Unlike conventional hybrid models based on coupled mesh generation, HYSCH constructs the meshes for the two submodels independently, allowing their spatial resolutions to be specified separately. This decoupled formulation provides greater flexibility in the numerical discretization and facilitates a more effective balance between computational cost and solution accuracy. A series of numerical experiments is conducted to quantify the sensitivity of the simulation results and convergence behavior to key numerical parameters and to assess **the numerical robustness and physical consistency** of the proposed model. The remainder of this paper is organized as follows. Section 2 describes the governing models, numerical algorithms, and coupling procedure for the heavy-species and electron submodels. Section 3 presents and discusses the numerical results used to evaluate the proposed model. Finally, Section 4 summarizes the principal conclusions of this study.

## 2. Method and Model

During Hall thruster operation, electrons are strongly magnetized and ionize the propellant neutrals in the discharge channel, while the generated ions are accelerated by the axial electric field to produce thrust. Electron transport toward the anode is sustained by electron–neutral collisions, electron–wall collisions, and anomalous transport induced by plasma instabilities. Inelastic interactions of electrons with neutrals and walls result in energy losses. In addition, ion–wall interactions can lead to recombination at the walls, whereas ion–neutral collisions result in momentum exchange.

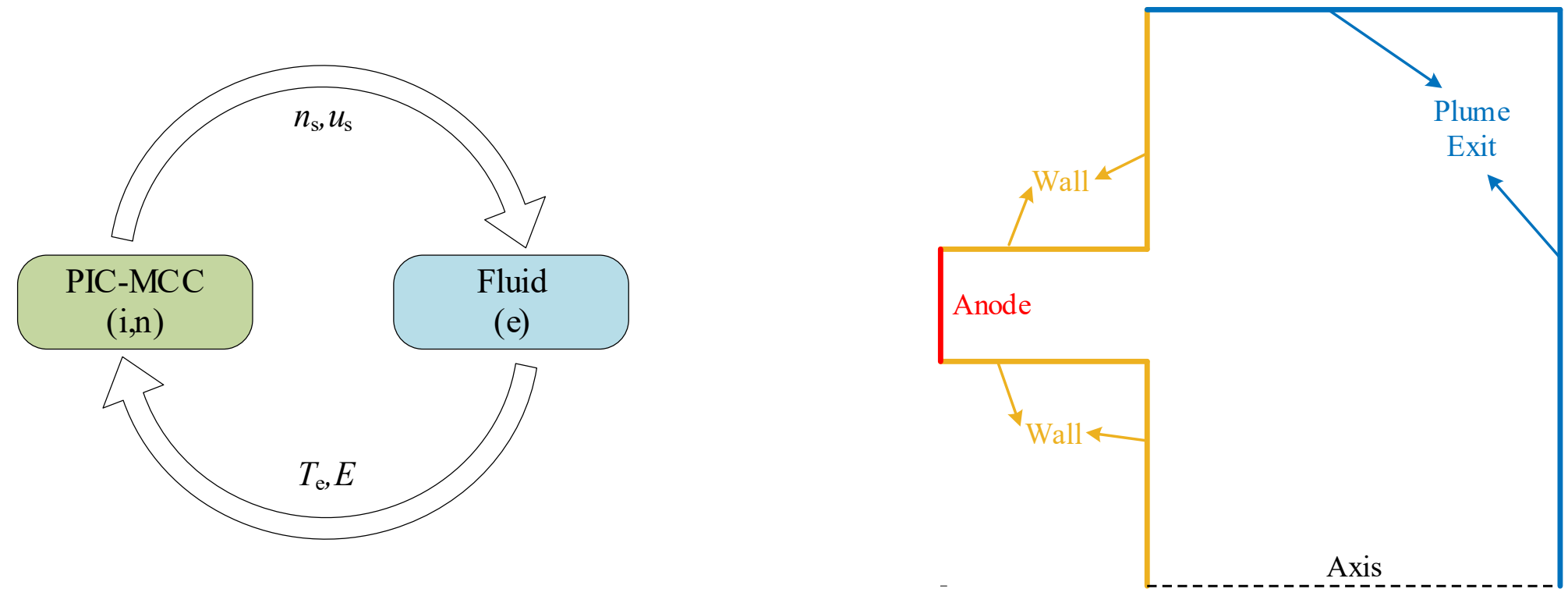

Figure 1 Schematic of the HYSCH framework  Figure 2 Computational region and boundary types

To model these processes, the present simulation framework consists of a heavy-species submodel and an electron submodel, as shown in Figure 1. The computational domain covers both the discharge channel and the plume region, with boundaries defined by the anode, channel walls, thruster symmetry axis, and plume outlet, as shown in Figure 2. The heavy-species submodel is solved using the particle-in-cell/Monte Carlo collision (PIC-MCC) method to capture the motion and collision behavior of heavy species in the electromagnetic field. The corresponding statistical moments, including heavy-species density and macroscopic velocity, are then supplied to the electron submodel. Assuming quasi-neutrality, the electron submodel treats electrons as a continuum and solves the fluid equations for the electron temperature and electric potential. These quantities are subsequently returned to the heavy-species submodel to update the motion and charge-state evolution of heavy species. The discharge process is obtained through repeated iterations of data exchange between the two coupled submodels.

### 2.1 The Heavy-Species Submodel

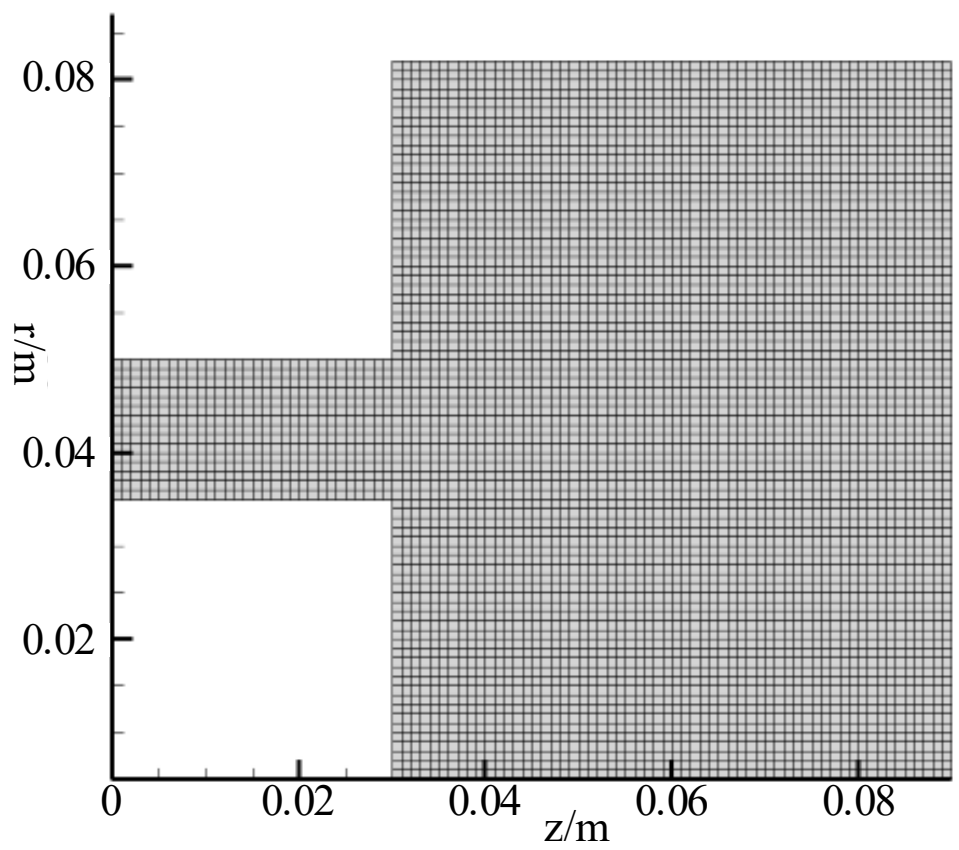

Figure 3 Uniform and structured mesh of heavy-species submodel

The heavy-species submodel employs a particle-in-cell/Monte Carlo collision (PIC-MCC) method on a uniform and structured mesh, as shown in Figure 3, to simulate the transport and collisional dynamics of ions and neutral species. Because a Hall-thruster discharge contains an extremely large number of physical particles, tracking every particle individually is computationally intractable. Instead, computational macroparticles are introduced, each representing $w_0$ physical particles. The macroparticle weight, $w_0$, is typically chosen within the range $10^8$–$10^{12}$, depending on the species and simulation conditions. The motion of each ion macroparticle is governed by Newton's second law subject to the Lorentz force, as shown in Equations (1) and (2), where $x$, $u$, $m$ and $q$ denote the ion position, velocity, mass and charge, respectively. The equations of motion are advanced using the Boris leapfrog scheme, which is widely used for charged-particle integration because of its long-term numerical stability and bounded energy error.

$$\vec{u} = \frac{\mathrm{d}\vec{x}}{\mathrm{d}t} \tag{1}$$

$$\vec{a} = \frac{\mathrm{d}\vec{u}}{\mathrm{d}t} = \frac{q}{m}\left(\vec{E} + \vec{u} \times \vec{B}\right) \tag{2}$$

Electron-impact ionization and ion–neutral charge-exchange (CEX) collisions are treated using the Monte Carlo collision (MCC) and direct simulation Monte Carlo (DSMC) methods, respectively. For a binary collision process, the expected number of collisions during a time step, $\Delta t$, depends on the relative speed of the collision partners, $g = |u_c - u_t|$, and the corresponding collision cross section, $\sigma_c(g)$. Assuming that collisions follow a Poisson process, the probability of at least one collision during a time step, $\Delta t$, is shown in Equation (3), where $\nu_c$ is the collision frequency. When $\nu_c\Delta t << 1$, this probability reduces to $P_c \approx \nu_c\Delta t$. The present model considers only electron-impact ionization and ion–neutral CEX collisions. Because these exhibit substantially different dependencies on relative speed and collision energy, as well as distinct cross-section characteristics, their numerical treatments are described separately below.

$$P_c = 1 - \exp(-\nu_c \Delta t) \tag{3}$$

Electron-impact ionization is modeled by treating each neutral macroparticle as interacting with a strongly magnetized electron population characterized by the local electron temperature, $T_e$. Because the electron thermal speed is substantially greater than the neutral-particle speed, the neutral contribution to the electron–neutral relative speed is neglected. Assuming a Maxwellian electron velocity distribution, the ionization process is characterized by the rate coefficient, as shown in Equation (4).

$$k_{\mathrm{ion}}\left(T_{\mathrm{e}}\right) = \left\langle \sigma_{\mathrm{ion}} u_e \right\rangle \tag{4}$$

where $u_e$ and $\sigma_{ion}$ denote the electron speed and the electron-impact ionization cross section, respectively. The rate coefficient is evaluated using the fitted expression for electron-impact single ionization of xenon reported by Goebel et al.[31]. For a neutral macroparticle $\boldsymbol{p}$, the ionization frequency is shown in Equation (5), and the probability of ionization during a time step is shown in Equation (6), where $n_{e,p}$ and $T_{e,p}$ are the electron number density and temperature interpolated at the position of macroparticle $\boldsymbol{p}$, respectively. An ionization event is accepted when a uniformly distributed random number $R \in [0, 1)$ satisfies $R < P_{\mathrm{ion,p}}$. Upon ionization, the neutral macroparticle is converted into a singly charged ion macroparticle at the same position and with the same velocity; electron momentum transfer and ionization recoil are neglected.

$$\nu_{\mathrm{ion,p}} = n_{e,p} k_{\mathrm{ion}}\left(T_{\mathrm{e,p}}\right) \tag{5}$$

$$P_{\mathrm{ion}} = 1 - \exp\left(\nu_{\mathrm{ion}}\Delta t\right) \approx \nu_{\mathrm{ion}}\Delta t \tag{6}$$

CEX collisions are modeled as binary interactions between ions and neutral atoms, with the ions treated as projectile particles. For a candidate ion–neutral pair, the relative speed is defined as $g_{in} = |u_i - u_n|$, and the corresponding CEX cross section, $\sigma_{CEX}(g_{in})$, is evaluated using the variable-hard-sphere (VHS) model reported by Miller et al.[32]. Within each computational cell, a neutral collision partner is sampled for each ion macroparticle according to the pairwise collision propensities, $\sigma_{CEX}(g_{in})\, g_{in}$. The collision probability is then calculated from Equation (7), and a CEX event is accepted when a uniformly distributed random number, $R \in [0, 1)$, is smaller than the calculated probability. Upon collision, the ion and neutral macroparticles exchange their charge states without changing their positions or velocities.

$$P_{\mathrm{CEX}} = 1 - \exp(\nu_{\mathrm{CEX}} \Delta t) \simeq \sigma_{\mathrm{CEX}}\left(g_{\mathrm{in}}\right) g_{\mathrm{in}} \Delta t \tag{7}$$

Boundary interactions of heavy species are also incorporated into the model. Neutral macroparticles are injected through the anode boundary with velocities sampled from a half-range Maxwellian distribution characterized by the anode temperature, $T_a$, with a drift velocity, $u_d$. When a heavy-species macroparticle impinges on a solid wall, any incident ion is assumed to be fully neutralized, and the particle is diffusely re-emitted into the computational domain as a neutral macroparticle. The velocity of the re-emitted particle is sampled from a half-Maxwellian distribution characterized by the wall temperature, $T_w$. At the axis of symmetry, specular reflection is imposed by reversing the velocity component normal to the boundary while leaving the tangential components unchanged. Macroparticles exiting the computational domain through the plume outlet are removed from the simulation.

Following the advancement of the macroparticle trajectories and the collision-induced update of their charge states, the particle information is deposited onto the computational grid using the cloud-in-cell (CIC) method, as illustrated in Figure 4. The particle shape function is defined in Equation (8), and the corresponding particle-to-node weighting coefficients are evaluated according to Equation (9). The number density, $n_{jk,s}$, and particle-flux density, $\Gamma_{jk,s} = (nu)_{jk,s}$, of heavy species $s$ at node (j,k) are subsequently calculated using Equations (10) and (11), respectively, where $j$ and $k$ denote the logical-grid indices of the node in the axial and radial directions. The deposited quantities are normalized by the corresponding nodal control volume, $V_{jk}$, centered at node (j,k) and defined in Equation (12). These nodal heavy-species moments are then supplied to the electron submodel for the present coupled iteration.

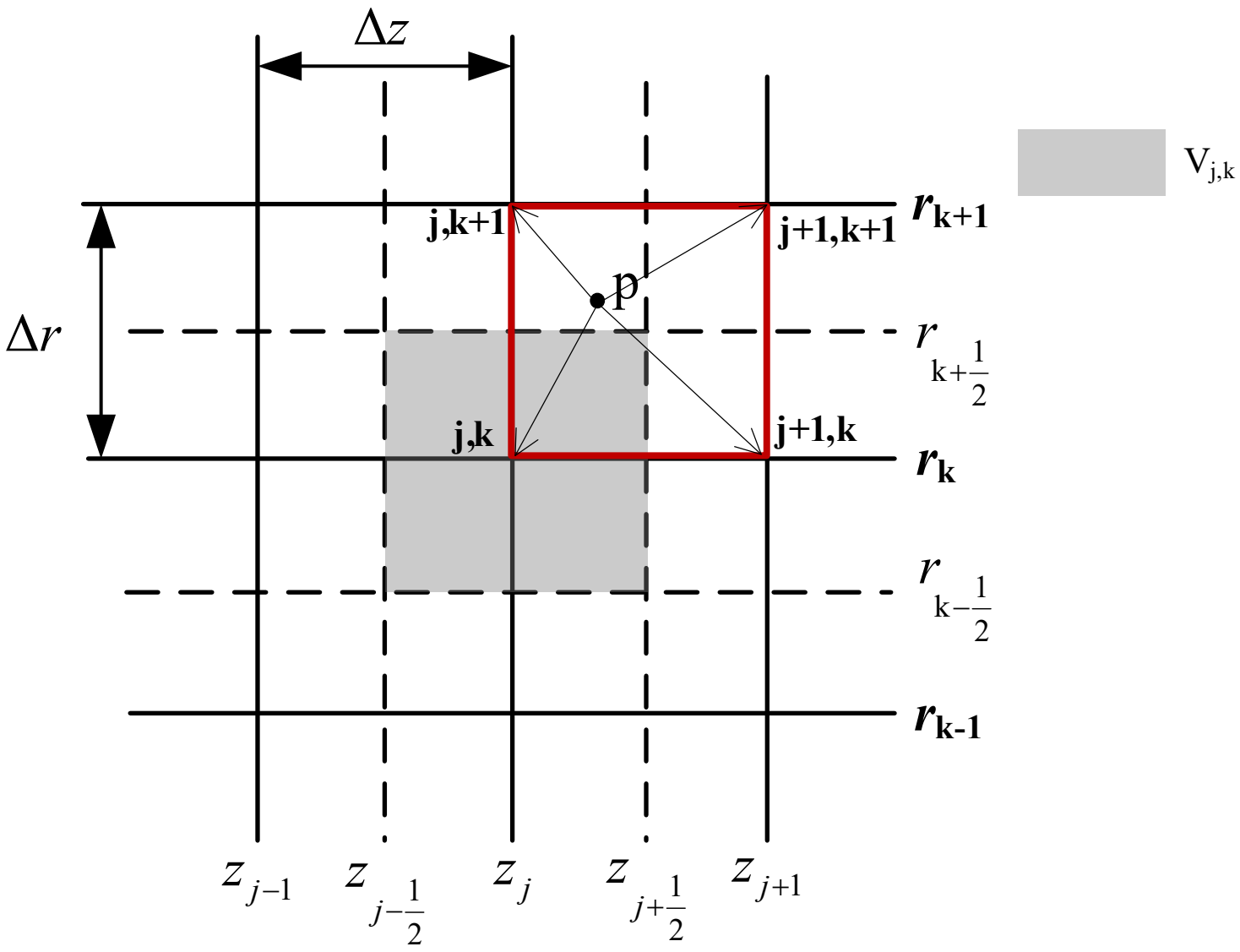

Figure 4 CIC particle deposition method

$$S_k=\begin{cases}1-\left|\dfrac{r_k-r_p}{\Delta r}\right|, & \left|\dfrac{r_k-r_p}{\Delta r}\right|<1\\ 0, & \left|\dfrac{r_k-r_p}{\Delta r}\right|<1\end{cases}$$
$$S_j=\begin{cases}1-\left|\dfrac{z_j-z_p}{\Delta z}\right|, & \left|\dfrac{z_j-z_p}{\Delta z}\right|<1\\ 0, & \left|\dfrac{z_j-z_p}{\Delta z}\right|<1\end{cases} \tag{8}$$

$$S_{j,k}=S_j\cdot S_k \tag{9}$$

$$n_{jk,\mathrm{s}}=\frac{\sum w_{0,\mathrm{s}}}{V_{\mathrm{jk}}} \tag{10}$$

$$\left(\overrightarrow{nu}\right)_{\mathrm{jk,s}}=\frac{\sum w_{0,\mathrm{s}}\vec{v}_{\mathrm{s}}}{V_{\mathrm{jk}}} \tag{11}$$

$$V_{\mathrm{jk}}=\pi\left(r^2_{k+\frac{1}{2}}-r^2_{k-\frac{1}{2}}\right)\Delta z \tag{12}$$

To correct the physical and numerical errors arising from finite-particle statistical sampling in the nodal control volumes near computational boundaries, particularly near the axis of symmetry, the correction procedure proposed by Verboncoeur et al.[33] is applied to the deposited heavy-species moments. In addition, the correct-weighting-boundary (CWB) method developed by Parra et al. [12] is employed to correct the systematic errors caused by one-sided particle-to-node deposition near the boundaries. At the sheath–presheath interface, the Bohm forcing condition (BFC) method[16] is applied to ensure the wall-directed normal component of the ion velocity satisfies the Bohm criterion. After these corrections, the corrected nodal number densities, $n_{\mathrm{jk,i}}$, and particle-flux densities $\Gamma_{jk,\mathrm{i}}$, are transferred to the electron submodel as the heavy-species input fields required to solve the current-continuity, electron-momentum, and electron-energy equations.

## 2.2 The Electron Submodel

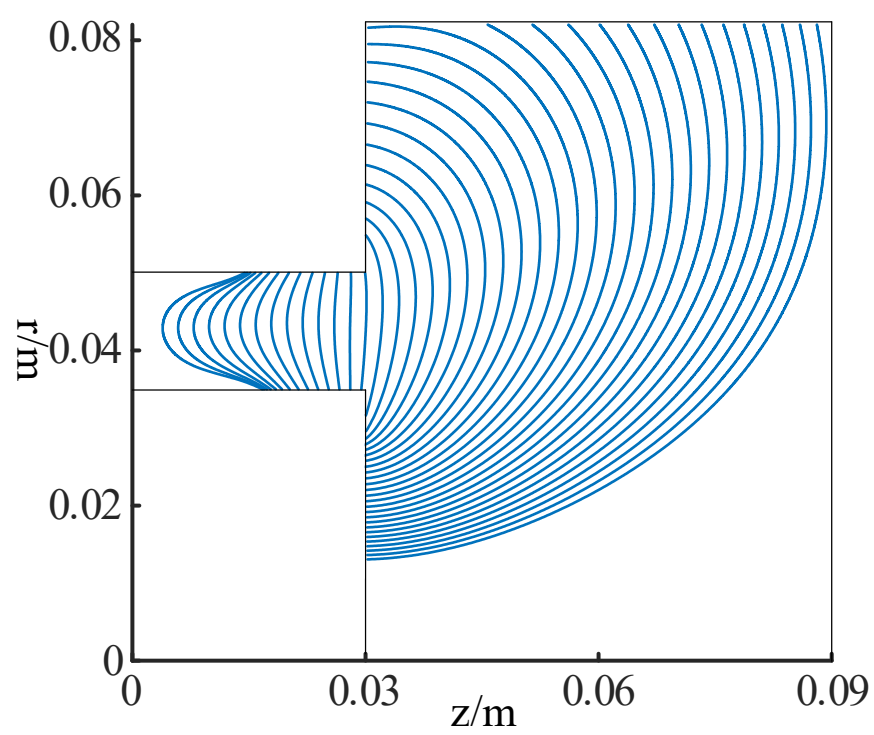

Figure 5 1D-MFAM of electron submodel

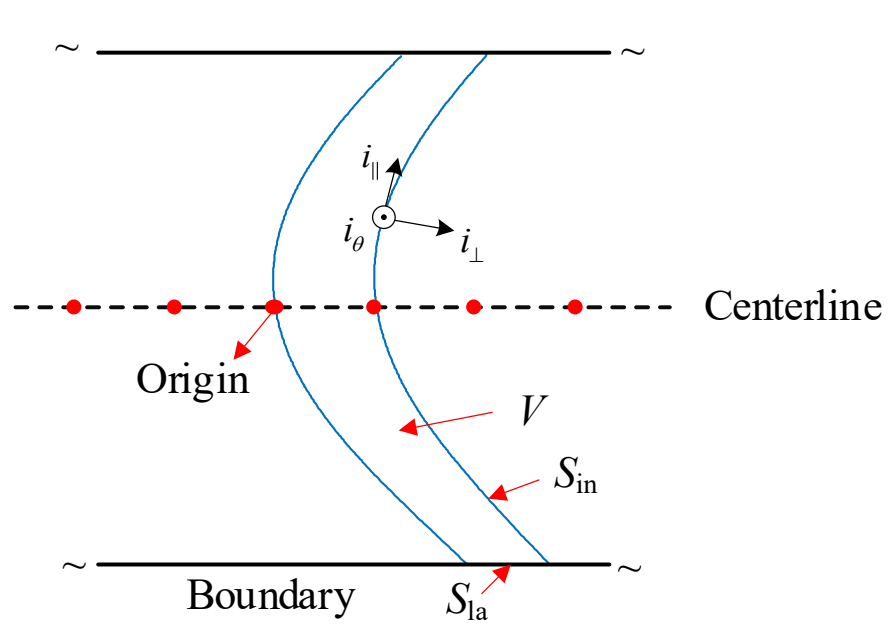

Figure 6 Generation of 1D-MFAM

The electron submodel is formulated using the thermalized-potential approach defined in Equation (13). Under this assumption, both the electron temperature $T_{\mathrm{e}}$ and the thermalized potential $\Phi^*$ are taken to be uniform along a given magnetic field line. The local potential, $\Phi$, at each position along a magnetic field line is evaluated from the thermalized potential $\Phi^*$, the electron temperature, $T_{\mathrm{e}}$, the local plasma density $n_e$, and the reference plasma density $n_e^*$, which is typically assumed to be constant. Electron mass and energy transport across the magnetic field are therefore described in the magnetic-field-aligned coordinate system $\{\hat{\iota}_{\perp}, \hat{\iota}_{\parallel}, \hat{\iota}_{\theta}\}$, with each magnetic field line treated as a computational unit. A 1D-MFAM is then constructed, as shown in Figure 5, using the streamline-tracing procedure illustrated in Figure 6. The magnetic field lines are traced from equally spaced points along the channel centerline with an interval of $\Delta x_e$. The resulting control volumes are bounded by adjacent magnetic field lines and physical boundaries, leading to a total number of $N_e$ control volumes. The inner and lateral surface areas of each control volume are denoted by $S_{in}$ and $S_{la}$, respectively, and its volume is denoted by $V$. The traced magnetic field lines are sequentially indexed from 1 to $N$, with the field line furthest upstream (index 1) defining the virtual anode boundary and that furthest downstream (index $N$) defining the virtual cathode boundary, where $N = N_e+1$ is the total number of traced magnetic field lines. Decoupling between the structured mesh and the 1D-MFAM is achieved by allowing $\Delta x_e \neq \Delta x$, which provides a higher level of flexibility in prescribing the spatial resolutions of the two submodels.

$$\phi = \phi^* + T_e \ln\left(\frac{n_e}{n_e^*}\right) \tag{13}$$

Furthermore, the magnetic stream function λ is introduced, as illustrated in Figure 7. The traced magnetic field lines are labeled by $\lambda_1$, $\lambda_2$, … $\lambda_N$, where $\lambda_1$ and $\lambda_N$ correspond to the virtual anode and virtual cathode boundaries, respectively. The relations among the electron temperature $T_e$, thermalized potential $\Phi^*$, and λ are given in Equation (14), while their gradients in the axisymmetric coordinate system are expressed in Equation (15). These relations enable a bidirectional transformation of plasma parameter descriptions between the axisymmetric coordinate and magnetic-field-aligned coordinate systems.

$$T_e = T_e(\lambda), \quad \phi^* = \phi^*(\lambda) \tag{14}$$

$$\nabla T_e = \frac{\partial T_e}{\partial \lambda}\nabla\lambda, \quad \nabla\phi^* = \frac{\partial \phi^*}{\partial \lambda}\nabla\lambda \tag{15}$$

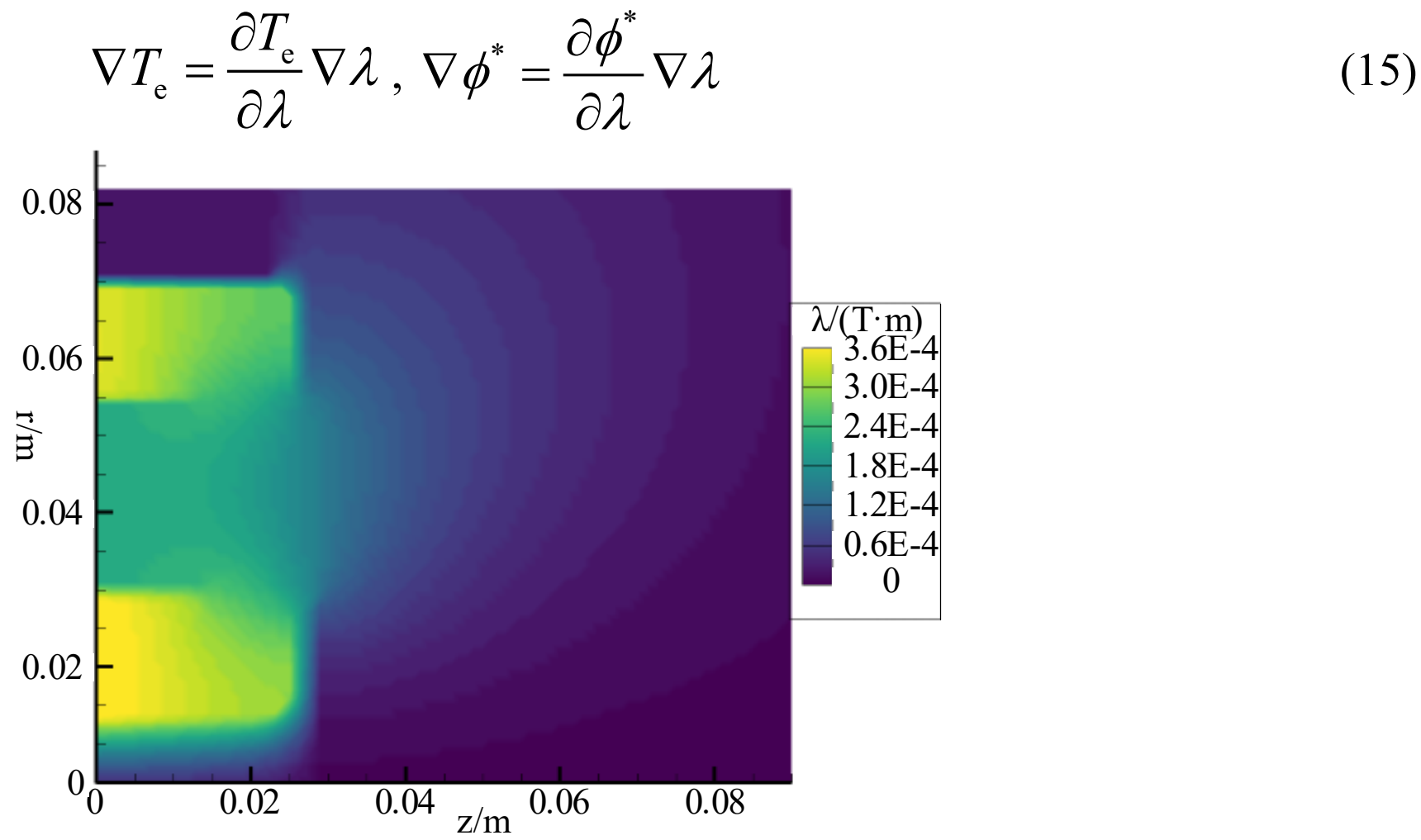

Figure 7 Distribution of magnetic stream function induced by Hall thruster

Subject to the quasi-neutrality condition as shown in Equation (16), the electron submodel consists of the current-, momentum-, and energy-conservation equations given in Equations (17)–(19), respectively. The heavy-species quantities required by the electron submodel are evaluated by interpolating the distributions supplied by the heavy-species submodel using the same weighting functions as those employed in CIC. The thermalized potential, $\phi^*(\lambda)$, is then determined by combining Equations (17) and (18), while the electron temperature $T_e(\lambda)$ is obtained by solving the energy-conservation equation, Equation (19).

$$n_e = n_i \tag{16}$$

$$I_{\mathrm{d}} = I_{\mathrm{e}} + I_{\mathrm{i}} = \int_{\lambda} \left( \overrightarrow{j_{\mathrm{e}\perp}} \cdot i_{\perp} \right) \mathrm{d}S + \int_{\lambda} \left( \overrightarrow{j_{\mathrm{i}\perp}} \right) \cdot i_{\perp} \mathrm{d}S \tag{17}$$

$$\frac{\partial m_{\mathrm{e}} n_{\mathrm{e}} \overrightarrow{u_{\mathrm{e}}}}{\partial t} + \nabla \cdot \left( m_{\mathrm{e}} n_{\mathrm{e}} \overrightarrow{u_{\mathrm{e}}} \overrightarrow{u_{\mathrm{e}}} \right) = -e n_{\mathrm{e}} \vec{E} - \nabla p + \vec{j}_{\mathrm{e}} \times \vec{B} - n_{\mathrm{e}} m_{\mathrm{e}} \nu_{\mathrm{e}} \overrightarrow{u_{\mathrm{e}}} \tag{18}$$

$$\frac{\partial}{\partial t} \left( \frac{3}{2} n_{\mathrm{e}} T_{\mathrm{e}} \right) + \nabla \cdot \left( \frac{5}{2} n_{\mathrm{e}} T_{\mathrm{e}} \overrightarrow{u_{\mathrm{e}}} + \overrightarrow{q_{\mathrm{e}}} \right) = -\vec{j}_{\mathrm{e}} \cdot \nabla \phi - Q_{\mathrm{inel}} \tag{19}$$

The current-continuity equation requires the total current integrated over any control surface associated with a magnetic field line spanning the computational domain to remain constant; the sum of the integrated electron and ion currents, $I_{\mathrm{e}}$ and $I_{\mathrm{i}}$, equals the discharge current. The electron momentum equation accounts for the electric and magnetic forces, the electron pressure-gradient force, and collisional friction, while electron inertia is neglected. Then, drift-diffusion approximations (DDA) are derived from the momentum-conservation equation, and the cross-field electron current density, $j_{\mathrm{e}\perp}$, and the cross-field electron conductivity, $\sigma_{\perp}$, are given in Equations (20) and (21), respectively. The condition $\Omega = \omega_{\mathrm{ce}}/\nu_{\mathrm{e}} >> 1$, where $\Omega$ is the Hall parameter, indicates that electrons are strongly magnetized. Here, $\omega_{\mathrm{ce}}$ denotes the electron cyclotron frequency, and $\nu_{\mathrm{e}}$ denotes the effective collision frequency governing electron transport. $\nu_{\mathrm{e}}$ is defined in Equation (22), where $\nu_{\mathrm{en}}$, $\nu_{\mathrm{ew}}$, and $\nu_{\mathrm{ano}}$ are the electron–neutral collision frequency, the equivalent collision frequency associated with near-wall conductivity, and the wave-induced anomalous collision frequency, respectively.

$$\overrightarrow{j_{e\perp}} = \frac{ne^2}{m_e \nu_{\mathrm{e}} (1+\Omega^2)} \left( \vec{E}_{\perp} - \frac{\nabla p_{\mathrm{e}\perp}}{n_{\mathrm{e}} e} \right) \tag{20}$$

$$\sigma_{\perp} = \frac{ne^2}{m \nu_{\mathrm{e}} (1+\Omega^2)} \tag{21}$$

$$\nu_{\mathrm{e}} = \nu_{\mathrm{en}} + \nu_{\mathrm{ew}} + \nu_{\mathrm{ano}} \tag{22}$$

The electron–neutral collision frequency, $\nu_{\mathrm{en}}$, comprises contributions from elastic scattering, excitation, and ionization, denoted by $\nu_{\mathrm{ela}}$, $\nu_{\mathrm{exa}}$ and $\nu_{\mathrm{ion}}$, respectively. The corresponding collision-rate coefficients are evaluated using the fitting formulae reported by Goebel et al.[31]. The equivalent collision frequency associated with near-wall conductivity, $\nu_{\mathrm{ew}}$, is evaluated using the quasineutral sheath model proposed by Ahedo et al. [34], as given in Equation (23). Here, the secondary-electron-emission (SEE) yield, $\delta_{\mathrm{w}}$, depends on the electron temperature and is defined as the ratio of the secondary-electron flux emitted from the wall to the incident primary-electron flux. For BN-$SiO_2$ walls, it is modeled as $\delta_{\mathrm{w}}(T_{\mathrm{e}}) = \min\{(T_{\mathrm{e}}/T_1)^{0.576}, 0.983\}$, where $T_1$ = 26.4 eV. The anomalous collision frequency $\nu_{\mathrm{ano}}$ is prescribed using the empirical model given in Equation (24), where the dimensionless empirical parameter $\alpha$ typically ranges from 0.3% to 1% within the discharge channel and from 3% to 8% in the plume region.

$$\nu_{\mathrm{ew}} = \frac{\delta_{\mathrm{w}}(T_{\mathrm{e}}) j_{\mathrm{ew}}}{1-\delta_{\mathrm{w}}(T_{\mathrm{e}})} \frac{S_{\mathrm{la}}}{V} = \frac{\delta_{\mathrm{w}}(T_{\mathrm{e}}) j_{\mathrm{iw}}}{1-\delta_{\mathrm{w}}(T_{\mathrm{e}})} \frac{S_{\mathrm{la}}}{V} \tag{23}$$

$$\nu_{\mathrm{ano}} = \alpha \omega_{ce} \tag{24}$$

Substituting Equations (14) and (15) into Equation (20) recasts the drift-diffusion approximation in the form given by Equation (25). Combining this relation with the current-continuity equation yields the field-aligned gradient of the thermalized potential, $\frac{\partial \phi^*}{\partial \lambda}$, as given in Equation (26)[11]. The discharge current, $I_d$, is determined by integrating Equation (26) from the virtual anode ($\lambda_1$) to the virtual cathode ($\lambda_N$) and enforcing the Dirichlet boundary conditions of thermalized potential, $\Phi^*(\lambda_1)$ and $\Phi^*(\lambda_N)$, as shown in Equation (27), where $\Phi_a$ denotes the anode potential, prescribed as the sum of the applied voltage $U_0$ and the anode sheath potential drop, $\Delta\Phi_{anode}$, as shown in Equation (28). Once $I_d$ has been obtained, the thermalized-potential distribution, $\Phi^*(\lambda)$, is reconstructed by solving Equation (26).

$$\overrightarrow{j_{e\perp}} = \sigma_\perp \left( -\frac{\partial \phi^*}{\partial \lambda} \nabla \lambda \cdot n - \nabla_\perp \left[ T_e \ln \left( \frac{n_e}{n_e^*} \right) \right] - \frac{\nabla p_{e\perp}}{n_e e} \right) \tag{25}$$

$$\frac{\partial \phi^*}{\partial \lambda} = -I_d \frac{1}{\int_\lambda \sigma_\perp \left( \nabla \lambda \cdot i_\perp \right) dS} + \frac{\int_\lambda n_i \left( \overrightarrow{u_{i\perp}} \cdot i_\perp \right) dS}{\int_\lambda \sigma_\perp \left( \nabla \lambda \cdot i_\perp \right) dS} - \frac{\int_\lambda \sigma_\perp \nabla_\perp \left[ T_e \ln \left( \frac{n_e}{n_e^*} \right) \right] \cdot i_\perp dS}{\int_\lambda \sigma_\perp \left( \nabla \lambda \cdot i_\perp \right) dS} \tag{26}$$

$$\begin{aligned} \phi^*(\lambda_1) &= \phi_a - T_{e,a} \ln \left( \frac{n_{e,a}}{n_e^*} \right) \\ \phi^*(\lambda_N) &= \phi_c - T_{e,c} \ln \left( \frac{n_{e,c}}{n_e^*} \right) \end{aligned} \tag{27}$$

$$\phi_a = U_0 + \Delta \phi_{anode} = U_0 + \left( T_e \ln \frac{S_{in} e n_0 \sqrt{T_e / 2\pi m_e}}{I_d - I_i} \right)_{\lambda_1} \tag{28}$$

To close the electron submodel, the electron temperature, $T_e(\lambda)$, is determined from the energy-conservation equation given in Equation (19), which accounts for convective and conductive thermal-energy transport, Ohmic heating, and energy losses due to inelastic processes. The conductive electron heat flux, $\overrightarrow{q_e}$, driven by the electron-temperature gradient, is defined in Equation (29). As given in Equation (30), the total inelastic energy-loss term $Q_{inel}$, comprises contributions from inelastic electron–neutral collisions and electron energy deposition at the walls, $Q_{ew}$. The electron–neutral contribution $Q_{en,inel}$ is expressed as shown in Equation (31), where $\langle\sigma u_e\rangle_{ion}$ and $\langle\sigma u_e\rangle_{exa}$ denote the electron-impact ionization and excitation rate coefficients, respectively, and $E_{ion}$ = 12.13 eV, $E_{exa}$ = 8.32 eV are the corresponding threshold energies.

$$\overrightarrow{q_e} = \frac{5T_e}{2e^2} \sigma_\perp \nabla T_e \tag{29}$$

$$Q_{inel} = Q_{ion} + Q_{exa} + Q_{ew} \tag{30}$$

$$Q_{en,inel} = Q_{ion} + Q_{exa} = n_n n_e \left[ E_{ion} \langle \sigma u_e \rangle_{ion} + E_{exa} \langle \sigma u_e \rangle_{exa} \right] \tag{31}$$

The wall-deposition contribution to the total inelastic energy loss, $Q_{ew}$, is evaluated using the sheath model developed by Ahedo et al.[34]. At the sheath–presheath interface $Q$, primary electrons originating from the bulk plasma are assumed to follow a Maxwellian velocity distribution characterized by the primary-electron temperature, $T_p$. The primary electrons that enter the sheath traverse the sheath potential difference, $\Phi_{wQ}$, and transfer their remaining kinetic energy to the wall upon impact. The resulting secondary electrons are emitted from the wall and accelerated by the sheath electric field toward the bulk plasma. Accordingly, the sheath potential difference, $\Phi_{wQ}$, and the net electron energy loss at the wall, $Q_{ew}$, are evaluated using Equations (32) and (33).

$$\phi_{wQ} = T_p \left[ \ln \sqrt{\frac{m_i}{2\pi m_e}} + \ln(1 - \delta_w) + \ln \left( \frac{n_{pQ} \sqrt{T_p / m_i}}{n_{iQ} v_{iQ}} \right) \right] \tag{32}$$

$$Q_{\text{ew}} = \Gamma_{\text{eQ}}\left[\left(\frac{2T_{\text{p}}}{1-\delta_{\text{w}}\left(T_{\text{p}}\right)}\right)+e\phi_{\text{wQ}}\right]S_{\text{la}}, \Gamma_{eQ} = \Gamma_{iQ} = n_{iQ}\left(\overrightarrow{u_{iQ}}\cdot\overrightarrow{n_{la}}\right) \tag{33}$$

Substituting Equations (14), (15), (26), (29)–(33) into Equation (19) and applying the forward-time-centered-space (FTCS) finite-difference scheme yields the discretized energy conservation equation given in Equation (34). During each electron-submodel iteration, the local pseudo-time step, $\Delta t_{\text{e}}$, is typically chosen within the range $10^{-3}\Delta t$ to $10^{-2}\Delta t$, where $\Delta t$ denotes the time step used to advance the heavy-species submodel. A Dirichlet boundary condition prescribing a fixed electron energy, $\varepsilon_{\text{e}}(\lambda) = \varepsilon_{\text{c}}$, is imposed at the virtual cathode (index $N$), whereas a homogeneous Neumann boundary condition, $\left.\frac{\partial T_{\text{e}}^{k+1}}{\partial \lambda}\right|_{\lambda=\lambda_1} = 0$, is imposed at the virtual anode. The discretized equation is then solved to obtain the updated electron-temperature distribution, $T_{\text{e}}^{k+1}(\lambda)$, after which its field-aligned gradient is obtained using the weighted least-squares (WLSQ) method. The updated temperature and temperature gradient are subsequently fed back into the current-continuity equation to update the thermalized potential $\Phi^{*}(\lambda_1)$, and the discharge current, $I_{\text{d}}$, thereby advancing the coupled electron-fluid solution to the next iteration.

$$T_{\text{e}}^{k+1} = T_{\text{e}}^{k} + f\left(T_{\text{e}}^{k}, \phi^{*,k+1}, n_{\text{s}}, \overrightarrow{u_{\text{s}}}\right)\Delta t_{\text{e}} \tag{34}$$

Once the accumulated advancement of the electron submodel reaches the global time step, $\Delta t$, the updated distributions of electron temperature, electrostatic potential, and other electron-fluid quantities are transferred to the heavy-species submodel. The coupled simulation is then advanced to the next global time step. For each macroparticle **p**, the magnetic-stream-function value, $\lambda_{\text{p}}$, is first evaluated at the particle position and subsequently used to interpolate the electron temperature, $T_{\text{e,p}}$, and other local electron-fluid quantities from the field -line distributions. The local electric field, $\overrightarrow{E_{\text{p}}}$, is then evaluated using Equation (35). These interpolated quantities are then used to advance the trajectory and update the charge state of macroparticle **p**.

$$\begin{aligned}\overrightarrow{E_p} &= -\left(\nabla\phi\right)_p \\ &= -\nabla\left(\phi^{*} + T_e \ln\left(\frac{n_e}{n_e^{*}}\right)\right)_p \\ &= -\left[\left(\frac{\partial\phi^{*}}{\partial\lambda}\right)_{\text{p}} + \left(\frac{\partial T_e}{\partial\lambda}\right)_{\text{p}} \ln\left(\frac{n_{e,p}}{n_e^{*}}\right)\right]\nabla\lambda + T_{e,p}\nabla\ln\left(\frac{n_{e,p}}{n_e^{*}}\right)\end{aligned} \tag{35}$$

### 2.3 Computational Workflow of HYSCH

In summary, the computational workflow of HYSCH is illustrated in Figure 8, and consists of three stages: preprocessing, initialization, and the main iterative loop. During preprocessing, the required input data are specified, including the thruster operating conditions, computational-domain geometry, boundary conditions, and numerical parameters for the two submodels. Based on the domain geometry and the prescribed magnetic-field configuration, the uniform structured mesh for the heavy-species submodel and the 1D-MFAM for the electron submodel are generated independently.

In the initialization, the neutral, ion, and electron distributions are constructed. The initial neutral distribution is obtained by advancing the injected neutral macroparticles using the PIC procedure without plasma interactions until the neutral flux through the plume outlet reaches a statistically stationary value. The ion population is then initialized throughout the computational domain with a uniform number density and a Maxwellian velocity distribution. Finally, the electron submodel is initialized by prescribing a uniform electron-temperature field and evaluating the initial discharge current, thermalized-potential distribution, and electrostatic-potential distribution from the initialized heavy-species fields.

Following initialization, the coupled solution is advanced using a global time step, $\Delta t$, after which the resulting heavy-species moments are deposited onto the structured-grid nodes. The electron submodel is subsequently advanced using the local time step, $\Delta t_e$, requiring $\Delta t/\Delta t_e$ electron iterations per global step, typically $10^2$–$10^3$. In the heavy-species advancement, the macroparticle charge states and trajectories are updated, and the resulting number densities and particle-flux densities are deposited onto the grid. In the electron-submodel iterations, the discharge current, thermalized-potential distribution, and electron-temperature distribution are updated successively. Once the accumulated electron-submodel advancement reaches $\Delta t$, the updated electron-fluid quantities are transferred to the heavy-species submodel, and the coupled simulation proceeds to the next global step.

This procedure is repeated until the time-averaged discharge-performance metrics and plasma-parameter profiles become statistically stationary, indicating that the simulated discharge has reached a statistically steady state.

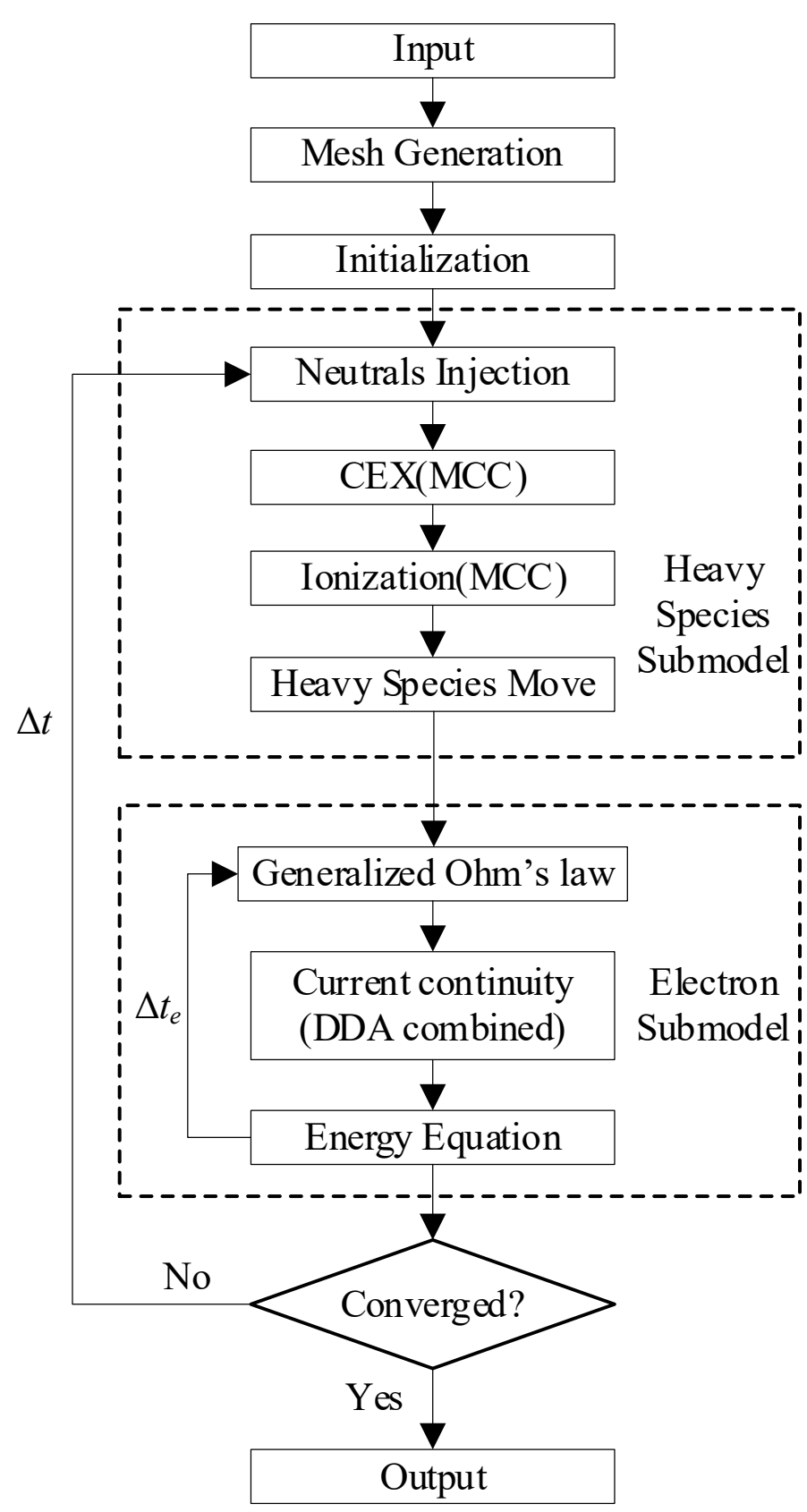

Figure 8 Computational workflow of HYSCH

## 2.4 Preliminary validation of HYSCH

To evaluate the performance of the HYSCH model, the discharge operation of a 1.35 kW Hall thruster is simulated. The thruster features an inner diameter of 70 mm, an outer diameter of 100 mm, and a channel length of 30 mm. The computational domain extends 90 mm in the axial direction and 77 mm radially. Xenon is used as the propellant, with an anode mass flow rate of 5 mg/s and an operating voltage of 300 V. The magnetic field configuration is similar to that of the SPT-100, which is commonly used in benchmark simulations. The magnetic field strength distribution along the discharge channel centerline is shown in Figure 9, with a peak value $B_{max}$ = 229.82 G.

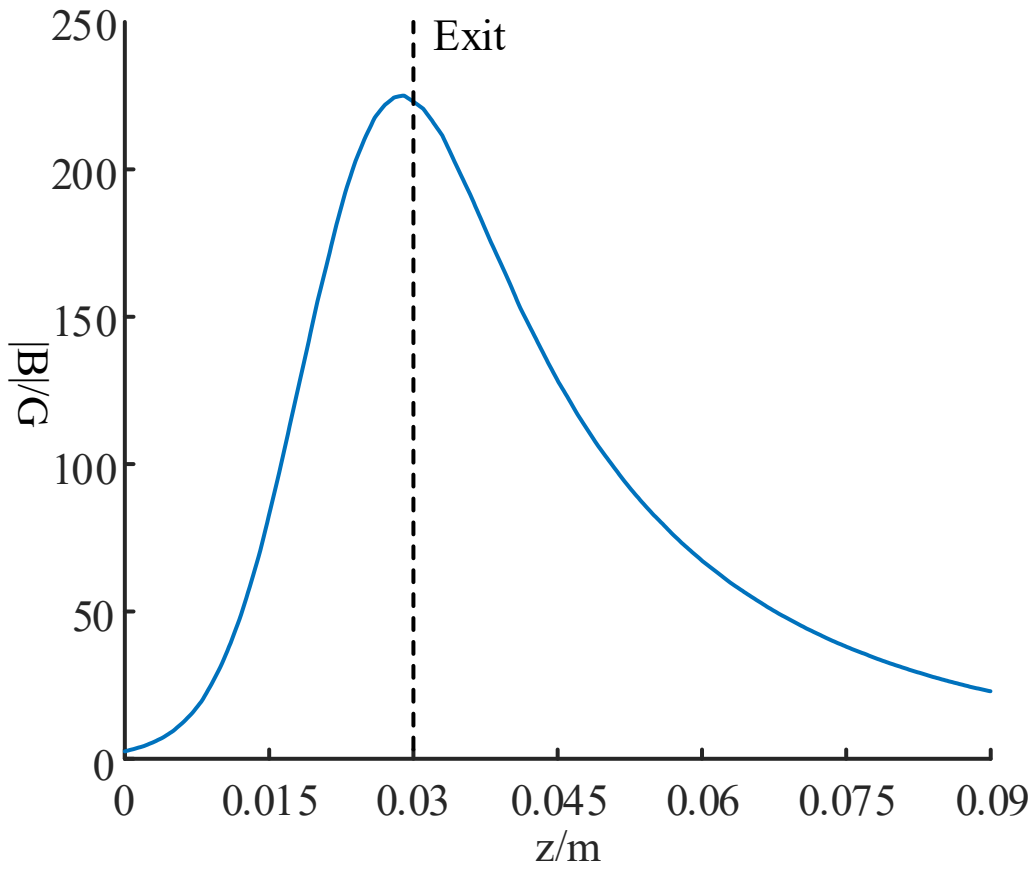

Figure 9 Distribution of magnetic field strength along the discharge channel centerline

Boundary conditions and numerical parameters are set as follows. The virtual anode tracing starts at an axial position of $z_a$ = 3 mm along the channel centerline, where the background magnetic field strength is 5.77 G (approximately 2.5% of $B_{max}$), which captures the weakly magnetized electron characteristics near the anode. The virtual cathode tracing starts at $z_c$ = 63 mm, with the traced magnetic field aligned with the actual cathode mounting position. The cathode electron energy is set to $\varepsilon_c = 1.5T_e = 5$ eV, consistent with the actual operating conditions. Empirical coefficients for wave-induced anomalous transport are set to 0.33% in the discharge channel and 6.25% in the plume region, consistent with those used in HPHall-2 by Hofer et al. [19]. For the electron submodel, $\Delta x_e$ = 1 mm, $\Delta t_e = 1.59\times10^{-10}$ s. For the heavy-species submodel, $\Delta x$ = 1 mm, $\Delta t = 200\Delta t_e$, $w_0 = 5.0\times10^8$. The displacement of ions, moving at their maximum velocity, within one time step is approximately 62.5% of the cell length, ensuring the Courant–Friedrichs–Lewy (CFL) criterion is satisfied.

The simulated performance parameters, summarized in Table 1, show good agreement with the measured data. Furthermore, time-averaged distributions of plasma density, potential, and electron temperature are compared with those obtained for the SPT-100 in HPHall-2[19], as illustrated in Figure 10. The distribution characteristics and numerical values show good consistency, confirming that HYSCH is fundamentally capable of simulating the discharge operation of Hall thrusters.

Table 1. Simulated performance parameters obtained with HYSCH

| Performance Parameters | Discharge current/A | Thrust/mN | Specific Impulse/s | Efficiency/% |
|---|---|---|---|---|
| Simulated | 3.94 | 80.92 | 1650 | 55.3 |
| Experimental | 3.98 | 79.62 | 1623 | 53.1 |
| Relative deviation/% | −1.01 | 1.63 | 1.63 | 4.14 |

To further validate the performance of HYSCH, a series of numerical experiments are conducted to assess its sensitivity to macroparticle weight, time step for heavy species, and spatial resolution of submodels and to confirm the model's robustness. The simulation case described above serves as the baseline. Additionally, the effect of virtual cathode and anode positions on the simulation results is analyzed to assess the model's physical consistency. All simulations reported in this work were performed on a server equipped with an AMD EPYC 9554 processor (64 cores, 128 threads, 3.1 GHz), with each case executed as a single-core serial job. To ensure a fair comparison of computational cost, all cases within a given parameter study were launched concurrently under identical conditions. Wall-clock times are therefore comparable within each study.

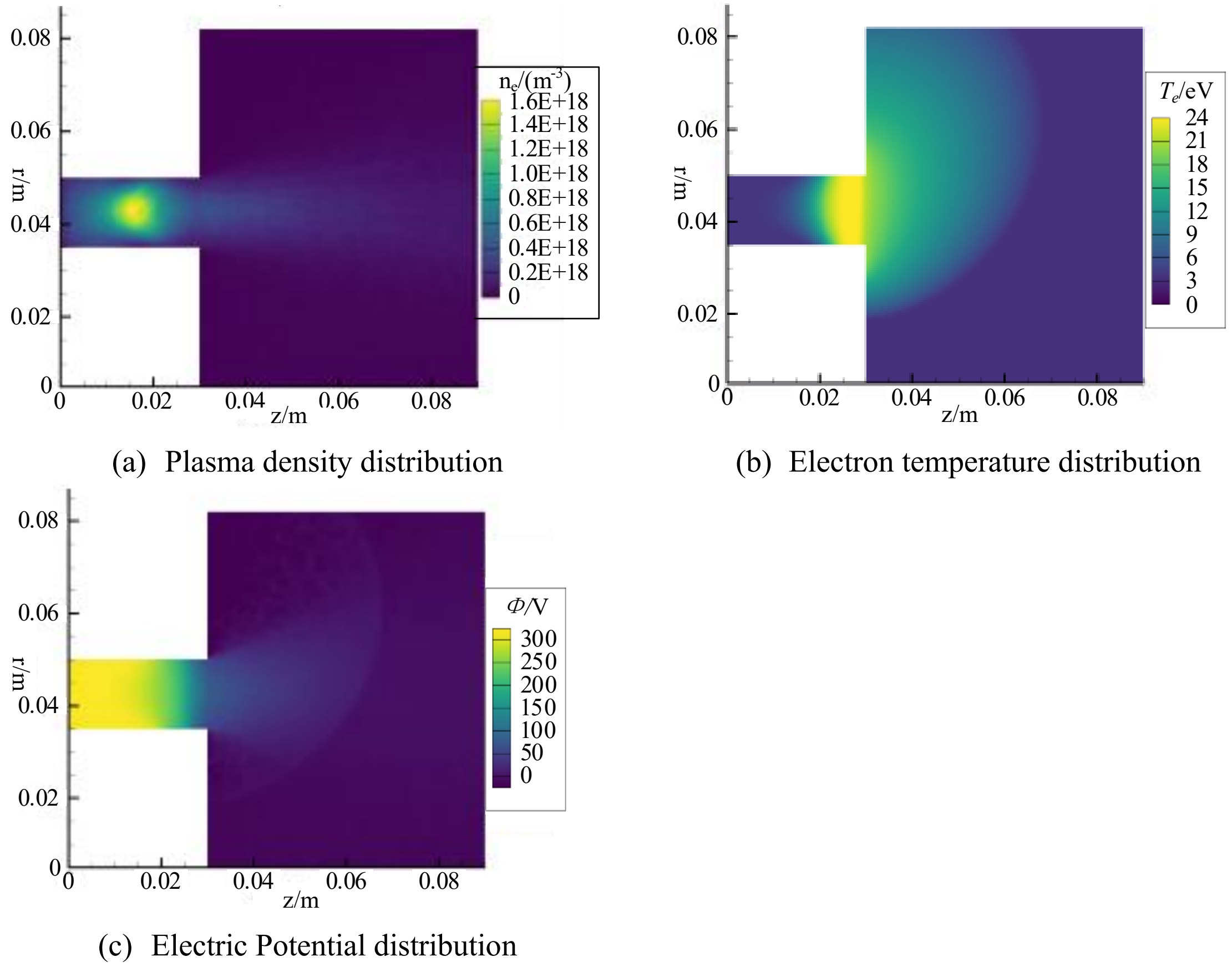

(a) Plasma density distribution

(b) Electron temperature distribution

(c) Electric Potential distribution

Figure 10 Plasma profiles simulated by HYSCH

## 3. Results and Discussion

3.1 Sensitivity to numerical parameters of the heavy-species submodel

With the physical particle population held fixed, the macroparticle weight, $w_0$, determines the number of computational macroparticles and therefore affects the statistical sampling error. In the baseline case, the time-averaged number of ion macroparticles is 273,657, corresponding to an average of approximately 54 ion macroparticles per grid cell. This sampling level is slightly above the commonly adopted empirical range of 30–50 macroparticles per cell. To investigate the sensitivity of the numerical solution to particle sampling, $w_0$ is varied over $2.5\times10^8$, $5.0\times10^8$, $1.0\times10^9$, and $2.0\times10^9$, corresponding to an average of approximately 108, 54, 27, and 13.5 macroparticles per grid cell, respectively. Figure 11 shows the simulated performance metrics obtained with these macroparticle weights. Over the tested range, the discharge current, thrust, specific impulse, and efficiency vary only marginally, and the predicted performance remains in good agreement with the measurements for every macroparticle weight considered. For the finest sampling level, $w_0 = 2.5\times10^8$, the relative deviations from the measured values, defined as (simulated − measured)/measured × 100%, are −0.92%, 1.63%, 1.63%, and 4.23% for the discharge current, thrust, specific impulse, and efficiency, respectively. Increasing the number of macroparticles therefore does not systematically improve the agreement with the measurements, whereas the computational cost increases substantially: the total wall-clock time rises from 7 h 56 min for $w_0 = 2.0\times10^9$ to 1 d 2 h 46 min for $w_0 = 2.5\times10^8$, a factor of approximately 3.4. These results indicate that the predicted global performance is largely insensitive to $w_0$ within the tested range, whereas the computational cost increases substantially, though sublinearly, with the number of macroparticles. The sublinear scaling reflects the fact that a considerable fraction of the total runtime is independent of the macroparticle population, being dominated instead by the electron-submodel iterations.

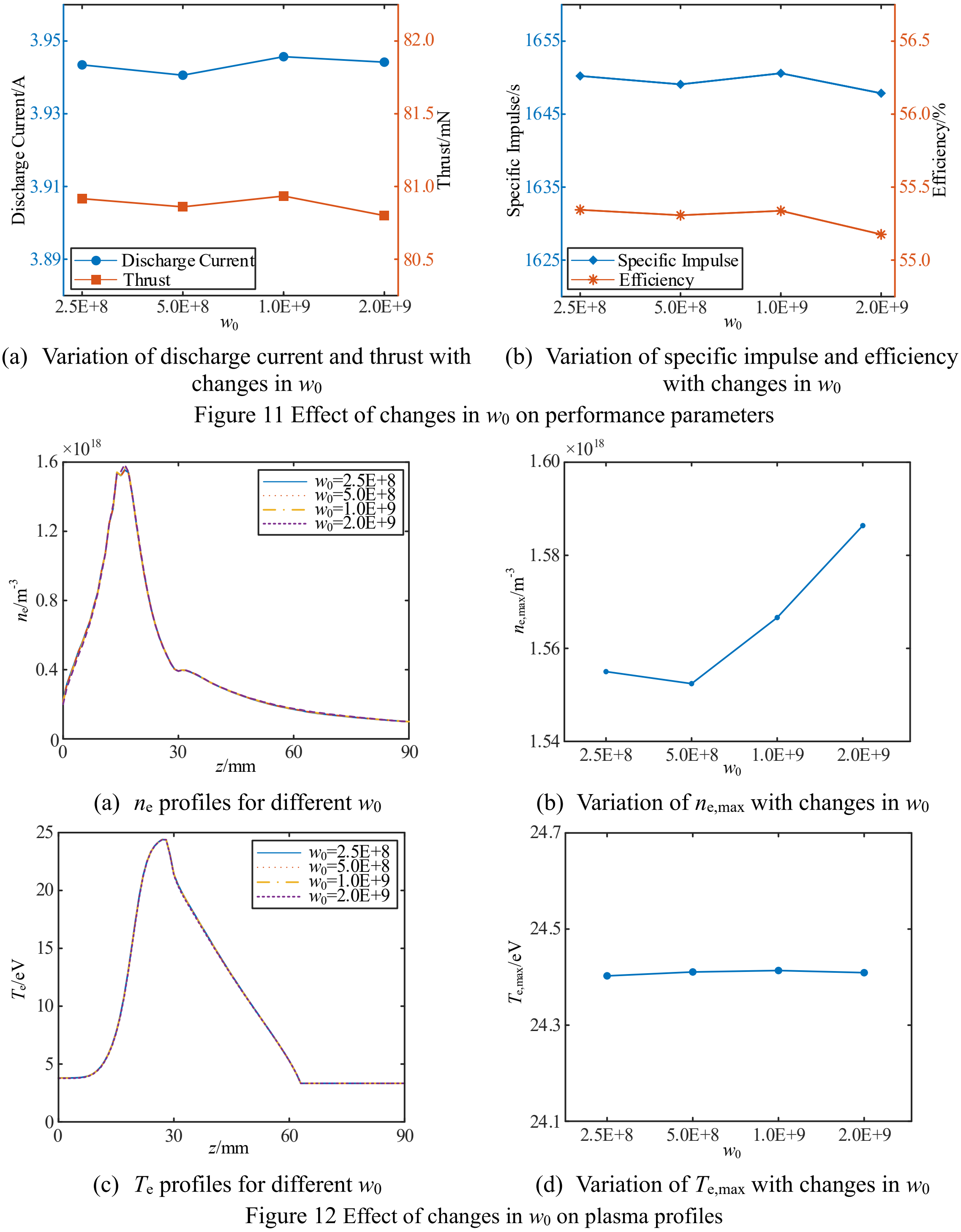

(a) Variation of discharge current and thrust with changes in $w_0$

(b) Variation of specific impulse and efficiency with changes in $w_0$

Figure 11 Effect of changes in $w_0$ on performance parameters

(a) $n_e$ profiles for different $w_0$

(b) Variation of $n_{e,max}$ with changes in $w_0$

(c) $T_e$ profiles for different $w_0$

(d) Variation of $T_{e,max}$ with changes in $w_0$

Figure 12 Effect of changes in $w_0$ on plasma profiles

Figure 12 shows the influence of $w_0$ on the time-averaged plasma-parameter profiles. The centerline profiles of plasma number density and electron temperature are selected for quantitative comparison. Figure 12 (a) and (b) present the centerline plasma number-density profiles and the corresponding peak values, respectively. As $w_0$ is reduced from $2.0\times10^9$ to $5.0\times10^8$, the peak density $n_{e,max}$ decreases by approximately 2.2%. A further reduction in $w_0$ to $2.5\times10^8$ produces no appreciable additional change, indicating that $n_{e,max}$ has approached a particle-number-converged value. The ionization region remains centered at an axial position of approximately 17 mm, with no discernible displacement, and the centerline number-density profiles exhibit no substantial differences elsewhere. The corresponding electron-temperature profiles are presented in Figure 12 (c) and (d). Overall, the time-averaged plasma-parameter profiles exhibit only weak sensitivity to the macroparticle weight as the number of computational particles is increased.

The heavy-species time step, $\Delta t$, determines the temporal discretization of the particle equations of motion and directly affects the accuracy with which macroparticle trajectories are integrated. Reducing $\Delta t$ generally decreases the temporal truncation error and improves the resolution of heavy-species transport. In the present analysis, the ratio $\Delta t/\Delta t_e$ is varied over 10, 25, 50, 100, 200, and 400, while the electron-submodel time step, $\Delta t_e$, is held fixed. Figure 13 shows the resulting variation in the simulated performance metrics. The case with $\Delta t = 400\Delta t_e$ violates the CFL criterion and is intentionally included as an extreme case to assess the consequences of exceeding the prescribed stability limit. As $\Delta t$ is reduced from $400\Delta t_e$ to $10\Delta t_e$, the predicted performance metrics approach time-step-independent values. Among them, the discharge current shows the most pronounced variation, resulting in a similar shift in efficiency. For the finest case with $\Delta t/\Delta t_e = 10$, relative deviations from the measured values are 0.11%, 1.58%, 1.58%, and 3.02% for discharge current, thrust, specific impulse and efficiency, respectively. Compared with the baseline case, the deviation in efficiency is reduced from 4.14% to 3.02%, while the others are also closer to the experimental values. However, this improvement is obtained at a considerably higher cost: the total wall-clock time rises from 15 h 3 min for $\Delta t = 400\Delta t_e$ to 2 d 8 h 38 min for $\Delta t = 10\Delta t_e$.

For the baseline case, the efficiency exhibits the largest relative deviation among the examined performance metrics, approximately 1.11% relative to the time-step-converged value. This deviation is sufficiently small for reliable prediction of the overall discharge performance. By contrast, in the extreme case with $\Delta t = 400\Delta t_e$, the efficiency deviates from the converged value by approximately 2.66%, representing a relatively high discretization error compared with other cases. These results indicate that, within the tested range $\Delta t \leq 200\Delta t_e$, the simulated performance is only weakly sensitive to the heavy-species time step.

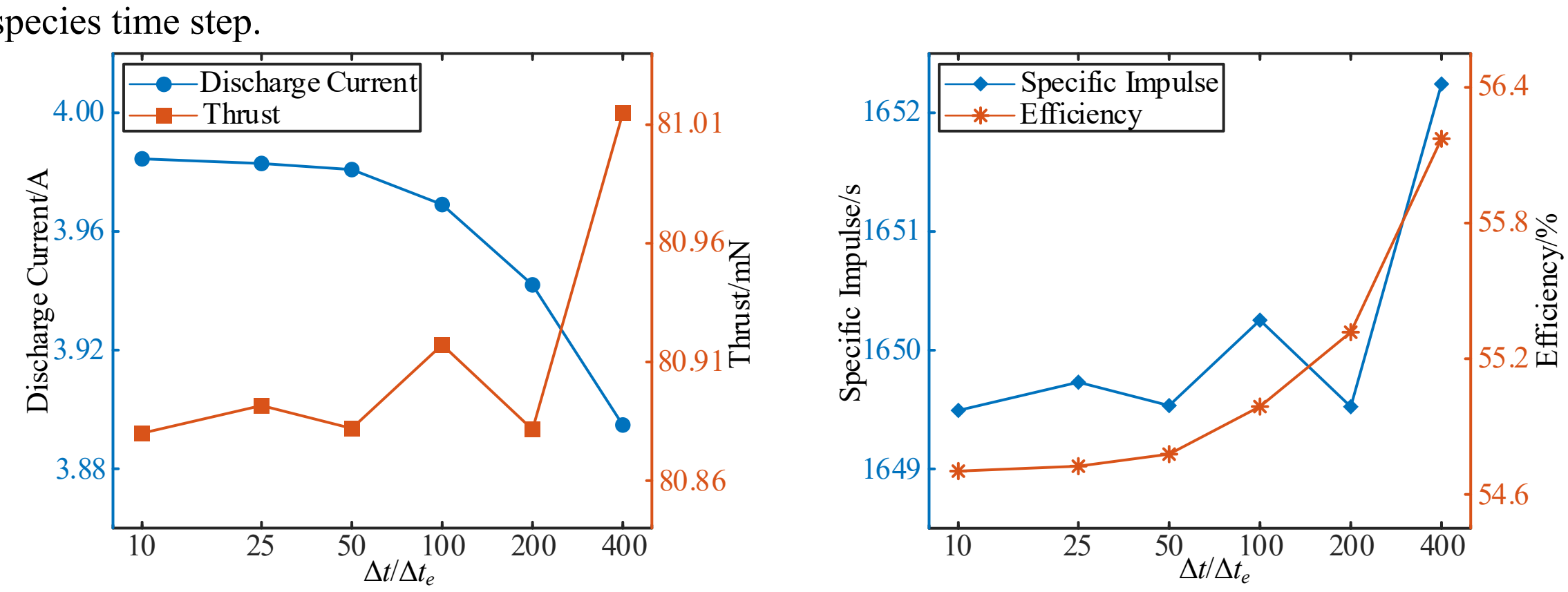

(a) Variation of discharge current and thrust with changes in $\Delta t$

(b) Variation of specific impulse and efficiency with changes in $\Delta t$

Figure 13 Effect of changes in $\Delta t$ on performance parameters

Figure 14 shows the influence of $\Delta t$ on the time-averaged plasma-parameter profiles. The centerline plasma number-density profiles and their peak values are presented in Figure 14 (a) and (b), respectively. No discernible displacement of the ionization region is observed as $\Delta t$ is varied. For $\Delta t \leq 200\Delta t_e$, $n_{e,max}$ remains within the narrow range $1.55\times10^{18}$ to $1.56\times10^{18}$ m$^{-3}$. When $\Delta t$ is increased to $400\Delta t_e$, however, $n_{e,max}$ increases to approximately $1.6\times10^{18}$ m$^{-3}$, representing a clear deviation from the time-step-converged range. In addition, the predicted plasma number density upstream of the ionization region is lower than that obtained in all the other cases. The centerline electron-temperature profiles and their peak values are shown in Figure 14 (c) and (d), respectively. The high-temperature region exhibits no appreciable axial displacement, and the overall variation in the peak electron temperature, $T_{e,max}$, remains small. Nevertheless, $T_{e,max}$ decreases monotonically as $\Delta t$ is reduced and gradually approaches a time-step-converged value. Taken together, these results demonstrate that the time-averaged plasma-parameter profiles are only weakly dependent on the heavy-species time step when $\Delta t \leq 200\Delta t_e$.

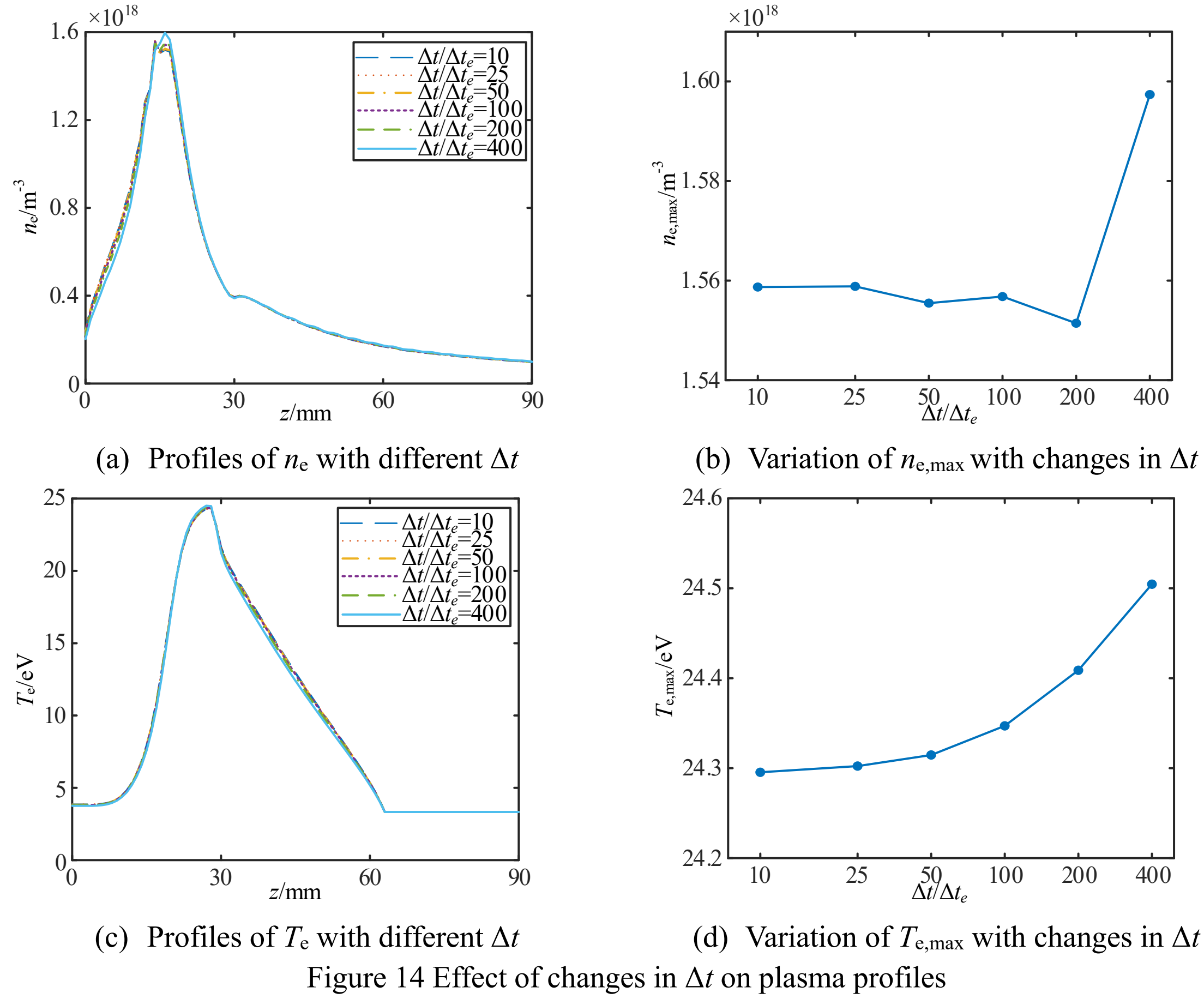

(a) Profiles of $n_e$ with different $\Delta t$

(b) Variation of $n_{e,max}$ with changes in $\Delta t$

(c) Profiles of $T_e$ with different $\Delta t$

(d) Variation of $T_{e,max}$ with changes in $\Delta t$

Figure 14 Effect of changes in $\Delta t$ on plasma profiles

These numerical experiments demonstrate the importance of satisfying the CFL criterion. Exceeding this limit primarily affects the acceleration region, where the large particle displacement over a single time step prevents the rapid ion transport and acceleration processes from being adequately resolved. This temporal under-resolution distorts the local plasma-density distribution, modifies the ionization dynamics, and ultimately causes deviations in both the time-averaged plasma profiles and the predicted performance.

### 3.2 Sensitivity to the spatial resolution of the two submodels

The decoupled-mesh formulation allows the spatial resolution of the two submodels to be varied independently. From a physical standpoint, this independence is justified because the characteristic length scales governing heavy-species and electron transport differ. Ion acceleration occurs over a distance determined by the axial electric field and the transit time, whereas electron energy transport is controlled by cross-field diffusion and the magnetic field topology. Numerical convergence studies must therefore assess whether each submodel can attain grid-independent solutions when refined separately, and whether the coupling procedure introduces spurious dependencies between the two resolutions. These questions are systematically examined as follows.

Although the heavy-species and electron submodels are discretized independently, spatial resolution is an important numerical parameter for both submodels and is characterized by the representative grid spacing and the corresponding number of grid cells. In the present numerical experiments, the electron-grid spacing, $\Delta x_e$, and the heavy-species grid spacing, $\Delta x$, are varied independently by adjusting $N_e$ and $N_{str,z}$, respectively, while holding the resolution of the other submodel fixed. Here, $N_e$ denotes the number of 1D-MFAM elements, whereas $N_{str,z}$ denotes the number of axial grid columns over the same axial extent in the heavy-species structured mesh. To maintain comparable numerical stability conditions during mesh refinement, the heavy-species and electron-submodel time steps, $\Delta t$ and $\Delta t_e$, are adjusted consistently with $\Delta x$ and $\Delta x_e$, respectively, to ensure fixed $\Delta x/\Delta t$ and $\Delta x_e/\Delta t_e$.

The sensitivity of the predicted performance to the spatial resolution of the electron submodel is examined first. The distance between the virtual anode and cathode is 60 mm. $\Delta x_e$ is set to 1.00, 0.75, 0.60 and 0.50 mm, corresponding to $N_e$ = 60, 80, 100 and 120. Figure 15 shows the variation in the predicted performance metrics as the electron mesh is progressively refined. All performance metrics decrease monotonically with increasing $N_e$, while the changes between successive refinements gradually diminish, indicating convergence toward grid-independent values.

Among all the examined metrics, the thrust exhibits the most pronounced variation, which in turn induces a relatively significant change in efficiency. For the case with the finest electron-mesh resolution, $N_e$ = 120, the relative deviations from the measured values are −1.98%, −0.26%, −0.26%, and 1.47% for the discharge current, thrust, specific impulse, and efficiency, respectively. Among all cases with varying $N_e$, the $N_e$ = 120 case yields the closest agreement with the measurements for thrust, specific impulse, and efficiency, indicating that refining the electron-submodel resolution improves the predictive accuracy. The associated cost, however, is substantial: the total wall-clock time increases from 17 h 6 min for the baseline case to 2 d 0 h 4 min, nearly tripling the computational cost.

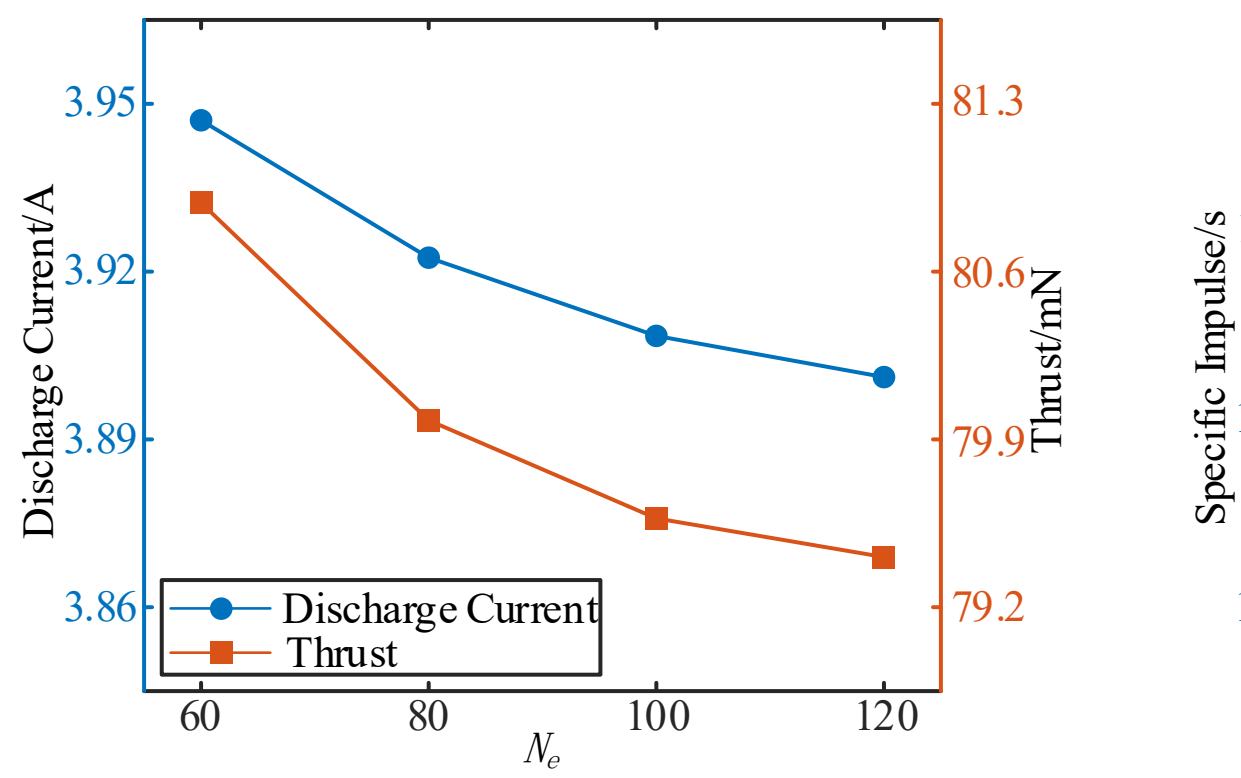

(a) Variation of discharge current and thrust with increasing $N_e$

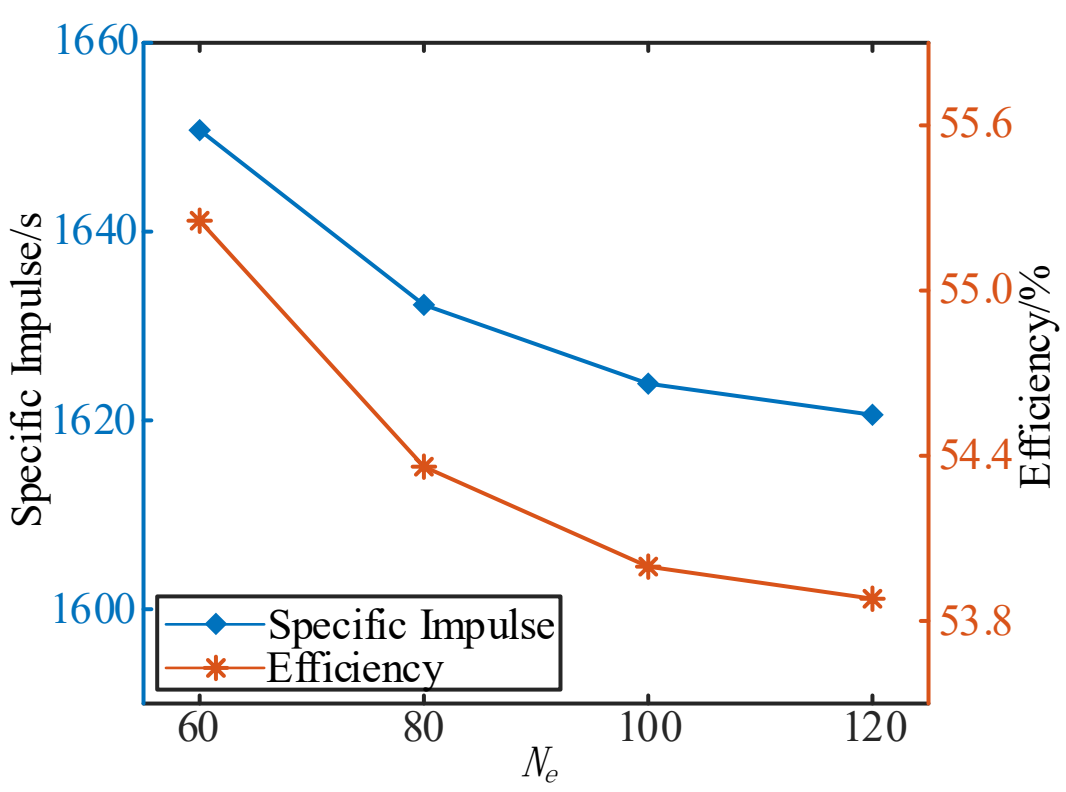

(b) Variation of specific impulse and efficiency with increasing $N_e$

Figure 15 Effect of increasing $N_e$ on performance parameters

Figure 16 shows the influence of the electron-submodel resolution on the time-averaged plasma-parameter profiles. The centerline plasma number-density profiles and their peak values are presented in Figure 16 (a) and (b), respectively. As the electron mesh is refined, the ionization region narrows slightly, while the peak electron number density, $n_{e,max}$, gradually approaches a grid-converged value. The maximum relative variation in $n_{e,max}$ is approximately 1.29%. The centerline electron-temperature profiles and their peak values are shown in Figure 16 (c) and (d), respectively. The high-electron-temperature region shifts slightly downstream with increasing $N_e$, and the peak electron temperature, $T_{e,max}$, decreases monotonically. These results indicate that the time-averaged plasma-parameter profiles tend to converge under successive mesh refinement. Because the changes become progressively smaller and the overall variation remains limited, the baseline case still reproduces the essential discharge characteristics.

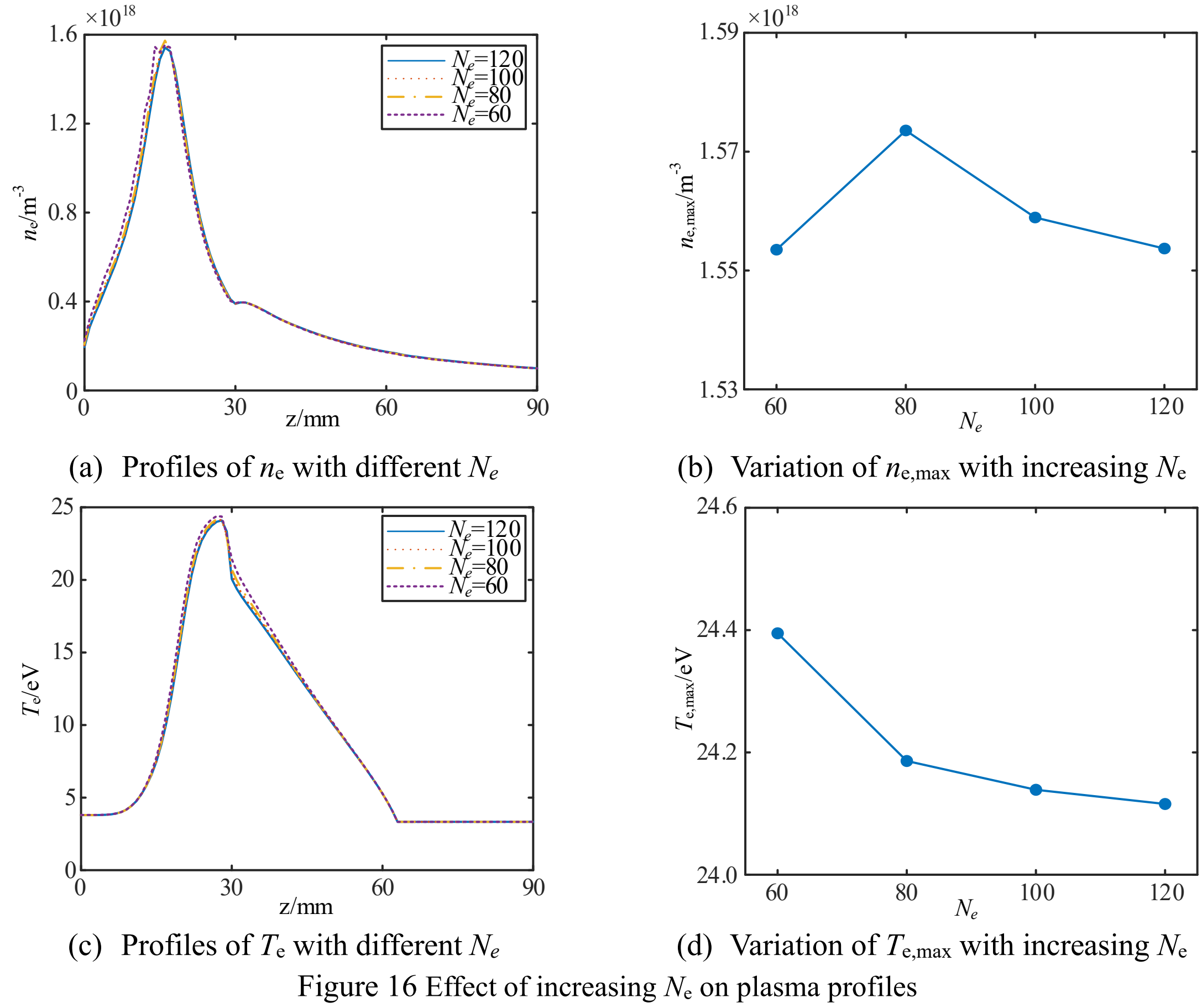

(a) Profiles of $n_e$ with different $N_e$

(b) Variation of $n_{e,max}$ with increasing $N_e$

(c) Profiles of $T_e$ with different $N_e$

(d) Variation of $T_{e,max}$ with increasing $N_e$

Figure 16 Effect of increasing $N_e$ on plasma profiles

The sensitivity of the numerical solution to the spatial resolution of the heavy-species submodel is examined next. The electron-submodel resolution is fixed at $N_e = 60$, while the heavy-species grid spacing, $\Delta x$, is varied by changing the number of axial grid cells, $N_{str,z}$, between the virtual anode and cathode. Specifically, $\Delta x$ is set to 1.00, 0.75, 0.60, and 0.50 mm, corresponding to $N_{str,z}$ = 60, 80, 100, and 120, respectively. Figure 17 presents the resulting variation in the predicted performance metrics. Although the performance metrics show a gradual convergence trend with increasing spatial resolution, the overall variations remain small and are notably less pronounced than those observed for reducing the heavy-species time step or increasing the electron-submodel resolution. For instance, the discharge current shows a total relative increase of approximately 0.88%, and the efficiency increases by only 0.40%, yielding no substantial improvement in agreement with the measurements. Meanwhile, the total wall-clock time rises from 17 h 6 min for $N_{str,z} = 60$ to 24 h 56 min for $N_{str,z} = 120$, an increase of approximately 46% in computational cost. The predicted performance is therefore only weakly dependent on the spatial resolution of the heavy-species submodel within the investigated range, and further refinement of $N_{str,z}$ offers limited benefit in the present configuration.

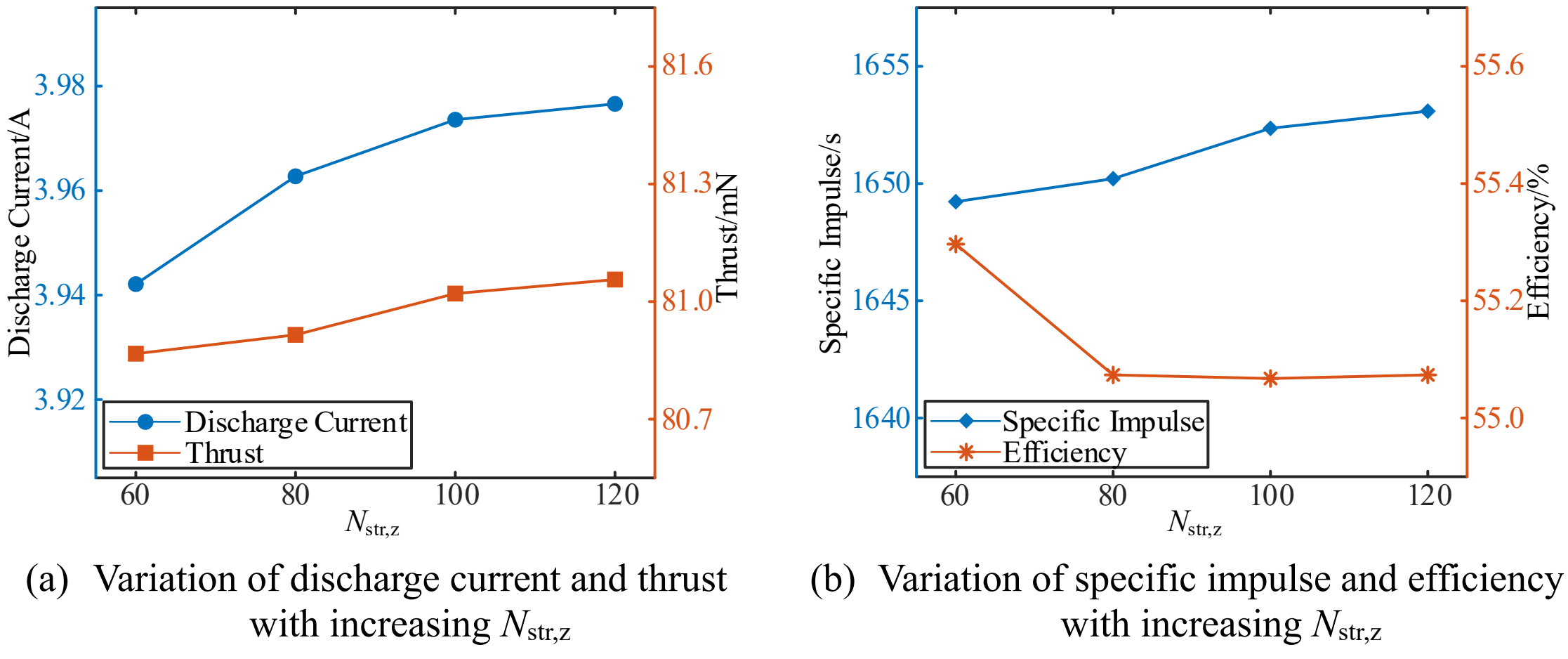

(a) Variation of discharge current and thrust with increasing $N_{str,z}$

(b) Variation of specific impulse and efficiency with increasing $N_{str,z}$

Figure 17 Effect of increasing $N_{str,z}$ on performance parameters

Figure 18 shows the effect of the heavy-species resolution on the time-averaged plasma-parameter profiles. The profiles obtained using the different meshes overlap closely over most of the computational domain, and no discernible displacement is observed in either the high-density or high-temperature region. As $N_{str,z}$ is increased from 60 to 120, corresponding to a twofold increase in axial resolution, the increments of both $n_{e,max}$ and $T_{e,max}$ diminish with successive refinement. The total relative variations in $n_{e,max}$ and $T_{e,max}$ are approximately 1.5% and 0.5%, respectively. These trends indicate convergence toward grid-independent profiles and demonstrate that the time-averaged plasma solution is only weakly sensitive to the heavy-species grid resolution over the tested range.

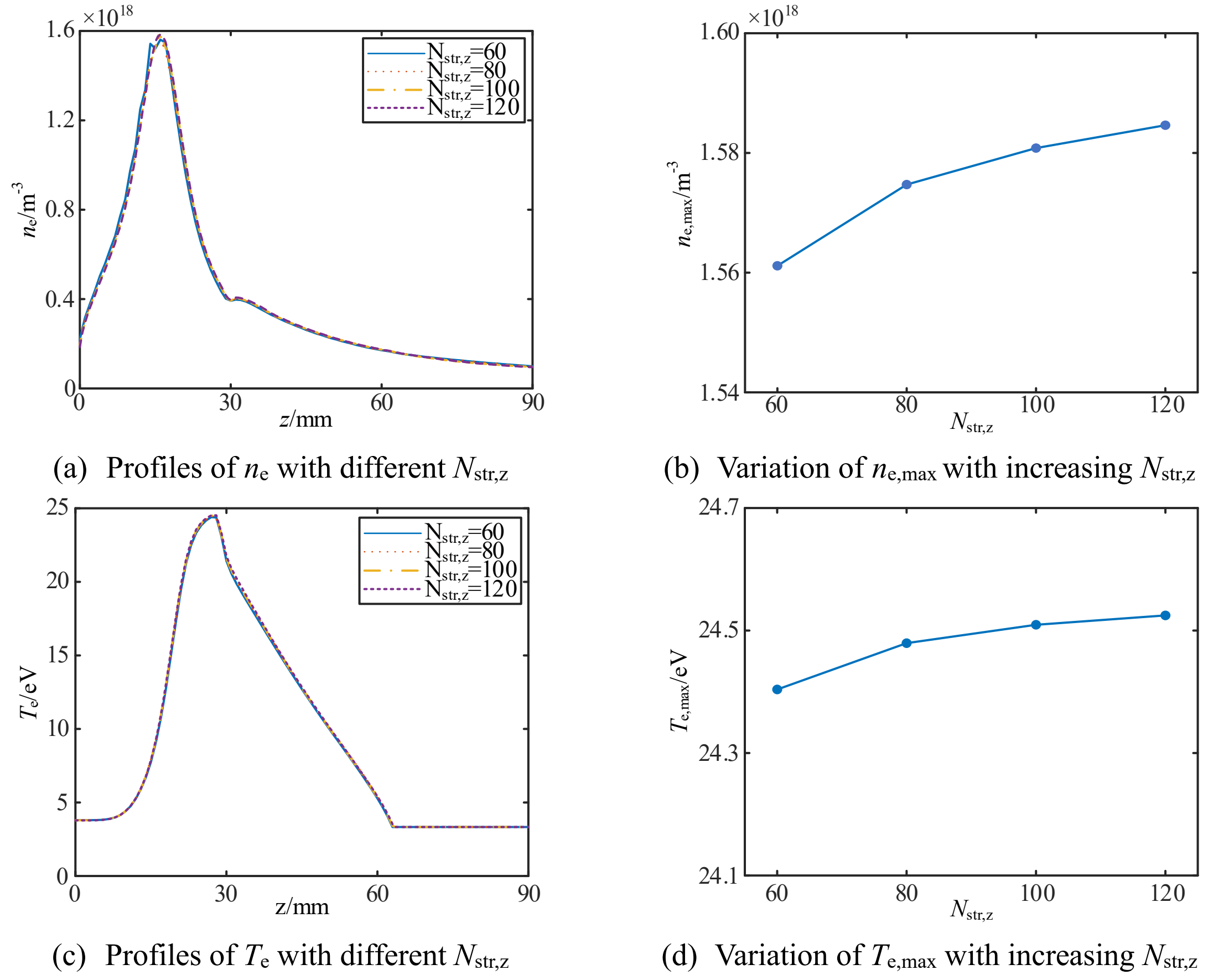

(a) Profiles of $n_e$ with different $N_{str,z}$

(b) Variation of $n_{e,max}$ with increasing $N_{str,z}$

(c) Profiles of $T_e$ with different $N_{str,z}$

(d) Variation of $T_{e,max}$ with increasing $N_{str,z}$

Figure 18 Effect of increasing $N_{str,z}$ on plasma profiles

The independent convergence observed under separate refinement of $N_e$ and $N_{str,z}$, together with the absence of systematic drift in the solution as either resolution is increased, confirms that the interpolation-based coupling between the two meshes does not introduce spurious resolution dependencies. The decoupled formulation therefore faithfully reproduces the physics of the coupled system while providing greater flexibility in balancing computational cost against accuracy.

Taken together, the preceding numerical experiments show that the HYSCH predictions exhibit only weak dependence or tendencies of convergence on the macroparticle weight, the heavy-species time step, and the spatial resolutions of the two submodels when these parameters are varied within the investigated ranges. These results confirm the **numerical robustness** of HYSCH. Importantly, the sensitivity analysis reveals that improving the temporal resolution of the heavy-species submodel (reducing $\Delta t$) and increasing the spatial resolution of the electron submodel (increasing $N_e$) yield more substantial improvements in agreement with the measurements, whereas increasing macroparticle number (reducing $w_0$) and refining the heavy-species spatial resolution (increasing $N_{str,z}$) produce only marginal gains. This asymmetry underscores the value of the decoupled-mesh formulation: because the two submodels exhibit distinct convergence characteristics, independent control of their resolutions enables more efficient allocation of computational resources than a coupled discretization would allow.

For the baseline numerical configuration, both the predicted performance metrics and the plasma-profile characteristics deviate from their respective numerically converged values by less than 2.5%. Because the changes become progressively smaller and the overall variation remains limited, the baseline case still reproduces the essential discharge characteristics while providing a favorable balance between accuracy and computational cost.

### 3.3 Validation of **physical consistency**

With all other numerical parameters fixed, the positions of the virtual cathode and anode are varied to evaluate the model's **physical consistency**. First, with $z_a$ and $\varepsilon_c$ fixed, $z_c$ is shifted downstream through 55 mm, 59 mm, 63 mm, 67 mm, 71 mm, 75 mm, and 79 mm, respectively. The variation in performance parameters is shown in Figure 19. As $z_c$ shifts downstream, discharge current, thrust, specific impulse and efficiency all exhibit strictly monotonic decreasing trends, with relative reductions of 1.26%, 1.98%, 1.98%, and 2.62%, respectively. Therefore, simulated performance parameters show a monotonic decrease with the downstream shift of the cathode position.

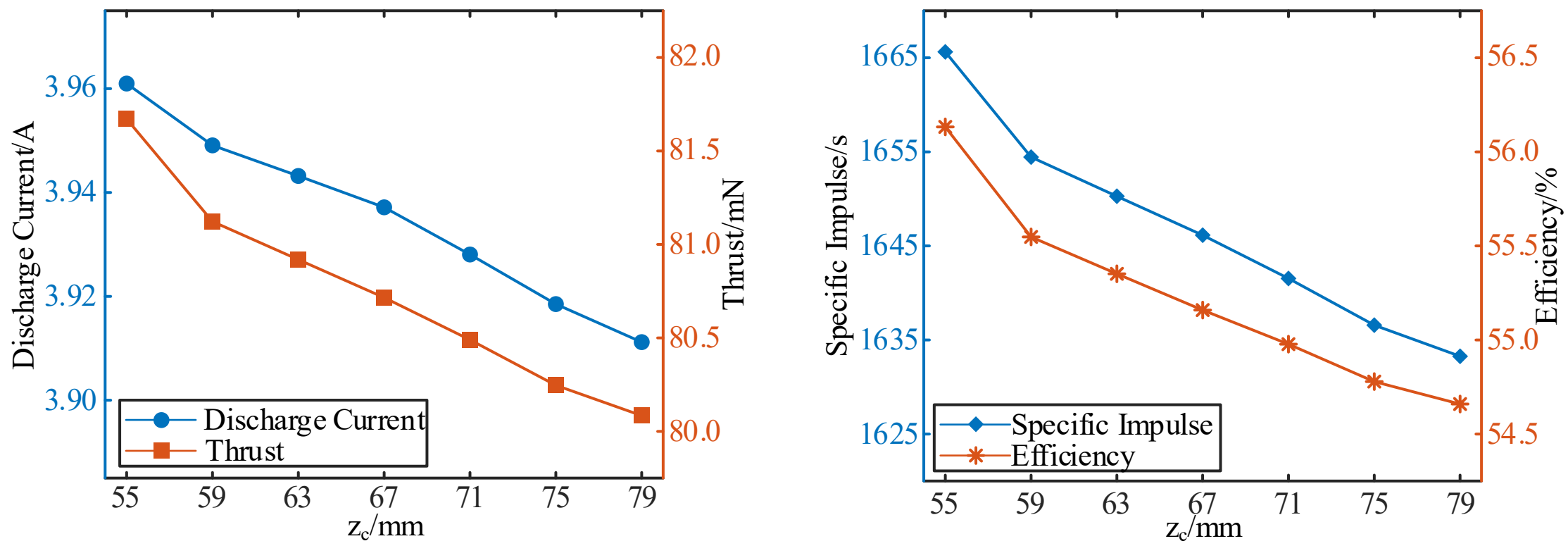

(a) Variation of discharge current and thrust with downstream shift of $z_c$

(b) Variation of specific impulse and efficiency with downstream shift of $z_c$

Figure 19 Effect of downstream shift of $z_c$ on performance parameters

The influence of cathode position on plasma profiles is shown in Figure 20. As $z_c$ moves downstream, $n_{e,max}$ decreases monotonically by $5.35\times10^{16}$ m$^{-3}$, corresponding to a relative variation of 3.39%, as shown in Figure 20 (b). Meanwhile, the ionization region contracts, and $n_e$ in the upstream region of the peak also shows a monotonic decrease, as shown in Figure 20 (a). $T_{e,max}$ decreases monotonically from 24.6 to 24.15 eV, with a relative variation of 1.83%, as shown in Figure 20 (d). Additionally, values of $T_e$ nearby also exhibit a monotonic decrease. However, since the profiles converge to the same value in the plume region, the temperature gradient decreases as the cathode shifts downstream.

The variation in plasma profiles is consistent with that of performance parameters, indicating that shifting the cathode downstream negatively affects the discharge operation in the HYSCH simulation. According to experimental studies by Yu et al. on cathode–thruster coupling[35], the underlying physical mechanism is as follows. Electrons emitted by the cathode need to be driven by the coupling voltage to enter the plume and channel. The downstream shift of the cathode increases the electron transport path and amplifies the coupling voltage loss, resulting in a reduction in effective voltage and a degradation in performance. This experimental conclusion aligns well with the results obtained from HYSCH.

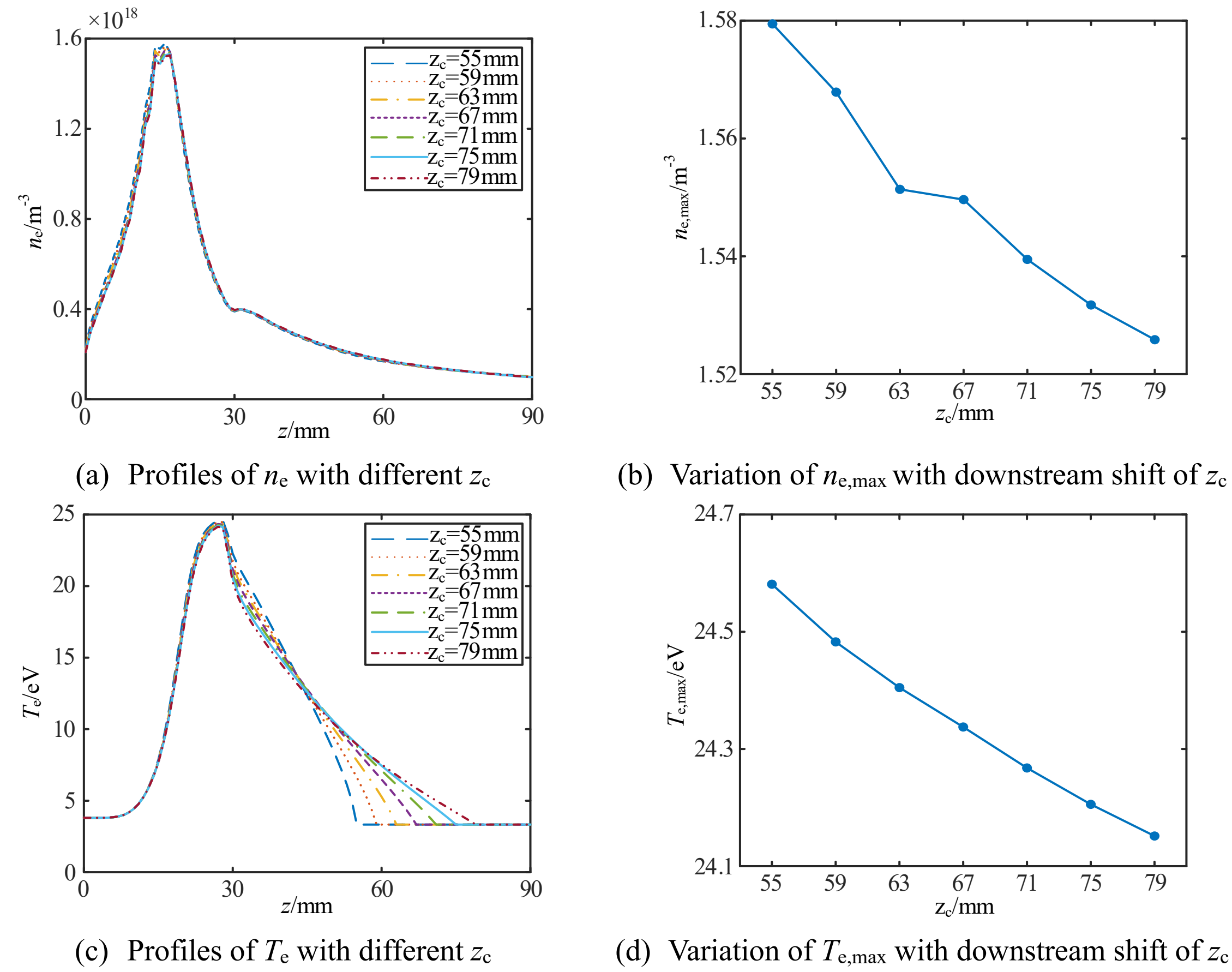

(a) Profiles of $n_e$ with different $z_c$ (b) Variation of $n_{e,max}$ with downstream shift of $z_c$

(c) Profiles of $T_e$ with different $z_c$ (d) Variation of $T_{e,max}$ with downstream shift of $z_c$

Figure 20 Effect of downstream shift of $z_c$ on plasma profiles

Furthermore, an equivalent circuit model is introduced to analyze the cathode–thruster coupling process. The total resistance between the anode and cathode consists of the anode-to-plume resistance $R_{ap}$ and the plume-to-cathode resistance $R_{cp}$. As the cathode moves downstream, $R_{cp}$ increases because the electron transport path is lengthened, resulting in a rise in coupling voltage. Correspondingly, the voltage across $R_{ap}$ decreases, leading to a reduction in the effective voltage. This, in turn, causes a degradation in performance. Similarly, a change of $R_{ap}$, induced by the shift of the anode, also affects discharge characteristics. The position $z_a$ is varied over 1, 2, 3, 4, 5, and 6 mm. The variation of performance parameters is shown in Figure 21, with all parameters exhibiting monotonic decreasing trends at an accelerating rate. The relative decreases in discharge current, thrust, specific impulse, and efficiency are 5.07%, 3.61%, 3.61%, and 1.19%, respectively.

The mechanism behind the performance variation is analyzed as follows. The cross-field conduction path of electrons in the channel is shortened, leading to a decrease in $R_{ap}$. Consequently, the effective voltage available for plasma discharge is reduced, resulting in degraded performance. This mechanism is identical to that of the cathode movement. In summary, the variation in performance with the change in anode position aligns with the conclusion and corollary of cathode–thruster coupling experiments, demonstrating that HYSCH is capable of capturing the underlying physical mechanisms.

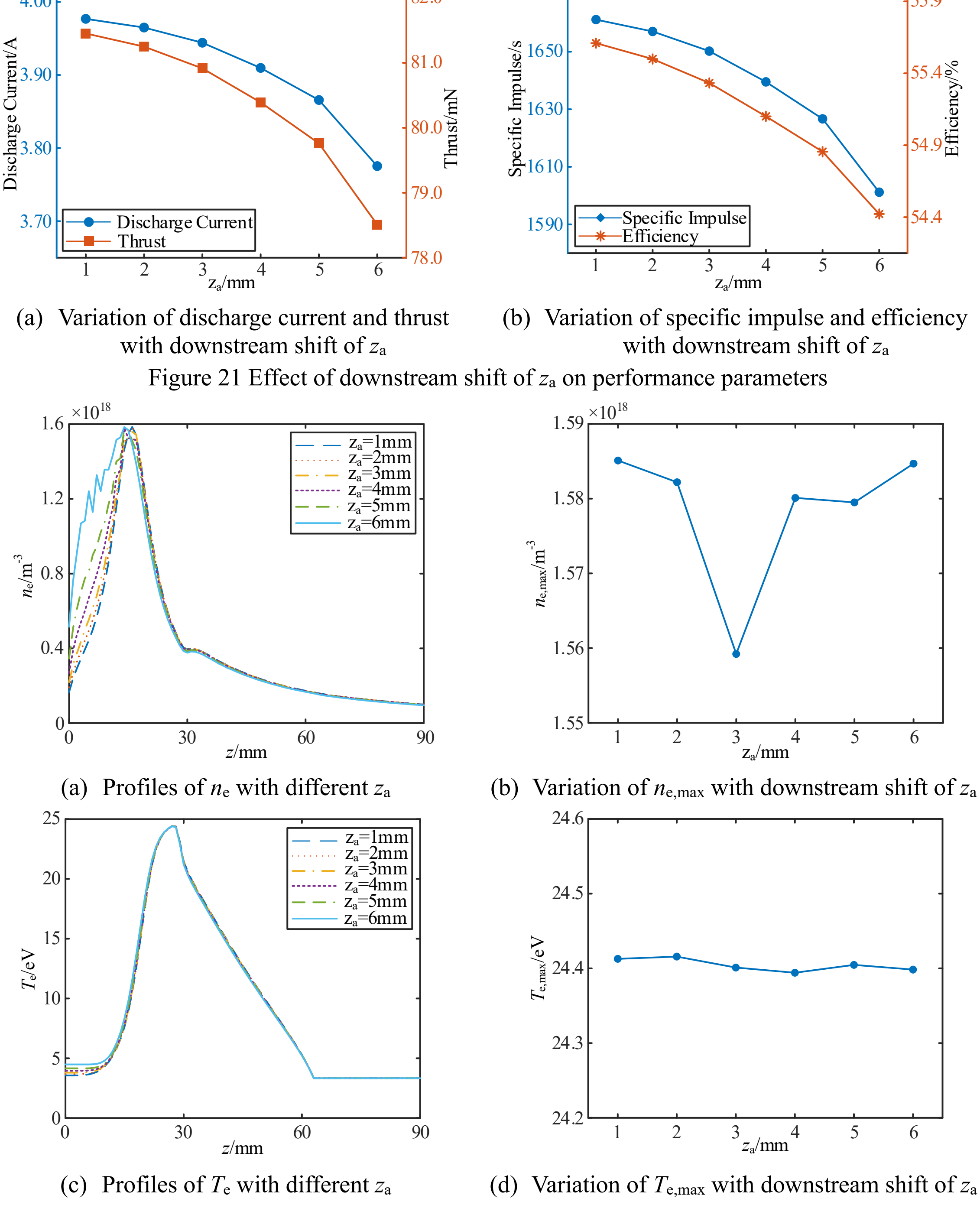

(a) Variation of discharge current and thrust with downstream shift of $z_a$

(b) Variation of specific impulse and efficiency with downstream shift of $z_a$

Figure 21 Effect of downstream shift of $z_a$ on performance parameters

(a) Profiles of $n_e$ with different $z_a$

(b) Variation of $n_{e,max}$ with downstream shift of $z_a$

(c) Profiles of $T_e$ with different $z_a$

(d) Variation of $T_{e,max}$ with downstream shift of $z_a$

Figure 22 Effect of downstream shift of $z_a$ on plasma profiles

The influence of anode position on plasma profiles is shown in Figure 22. As illustrated in Figure 22 (a) and (b), $n_{e,max}$ does not exhibit a monotonic variation trend with the movement of $z_a$. However, the ionization region gradually shifts and extends upstream, accompanied by monotonic increasing $n_e$ in the near-anode region. A distinct distortion occurs when $z_a$ = 6 mm. This phenomenon is linked to variations in the electron temperature profile, as shown in Figure 22 (c) and (d). Although $T_{e,max}$ does not change significantly, $T_e$ increases monotonically with the shift of $z_a$ in the near-anode region. This enhances the preionization upstream of the channel, causing the observed density distortion.

The underlying cause is as follows. As the virtual anode shifts downstream, the validity of the zero-temperature-gradient boundary condition is degraded in two aspects. First, the background magnetic field strength increases and the cross-field electron conductivity decreases, which gives rise to a non-negligible temperature gradient and thermal conduction. Second, the reduced conductivity enhances the local electric field and the associated Ohmic heating. The resulting energy increment must be balanced by net outflow and dissipation through thermal convection and inelastic collisions. The elevated electron temperature at the anode boundary therefore enhances numerical preionization upstream of the anode and leads to a distortion of the plasma density.

## 4. Conclusion

A hybrid simulation code for Hall thrusters, referred to as HYSCH, has been developed by coupling a kinetic heavy-species submodel with an electron-fluid submodel on fully decoupled meshes. The heavy-species submodel employs a uniform structured mesh and a particle-in-cell/Monte Carlo collision (PIC-MCC) method, whereas the electron submodel solves the current-continuity, electron momentum- and energy-conservation equations on a 1D-MFAM. Unlike conventional hybrid codes that enforce identical spatial discretizations for both submodels, HYSCH generates the two meshes independently and couples the two submodels through interpolation-based data transfer. This decoupled formulation exploits the distinct transport anisotropies of electrons and heavy species: electrons diffuse primarily along magnetic field lines, whereas ions are accelerated by the electric field with no preferential alignment. By allowing each submodel to adopt its own optimal resolution, the decoupled strategy provides greater flexibility in balancing computational cost and numerical accuracy and facilitates independent sensitivity studies.

Numerical sensitivity analyses were performed with respect to the macroparticle weight, heavy-species time step, and spatial resolutions of both submodels. Within reasonable parameter ranges, the predicted performance metrics and time-averaged plasma profiles exhibit tendencies of convergence or weak dependence on these numerical parameters, **confirming the numerical robustness of HYSCH.** Importantly, the sensitivity studies reveal an asymmetry in convergence behavior: refining the temporal resolution of the heavy-species submodel and the spatial resolution of the electron submodel yield more substantial improvements in predictive accuracy than reducing the macroparticle weight and refining the heavy-species spatial resolution. This asymmetry confirms that the two submodels have distinct resolution requirements, and that independent control of their discretizations—enabled by the decoupled-mesh formulation—permits more efficient allocation of computational resources than a coupled approach.

The predicted discharge performance deteriorates monotonically as the virtual cathode and anode boundaries are shifted downstream, in agreement with experimental observations and the expected behavior of cathode–thruster coupling. This agreement supports the **physical consistency** of the model. Nevertheless, a downstream displacement of the virtual anode increases the near-anode electron temperature and artificially enhances preionization, which can increase and eventually distort the predicted plasma number density. The virtual anode should therefore be located in a region with a sufficiently weak background magnetic field and high effective electron conductivity, where the imposed zero-temperature-gradient boundary condition remains a physically reasonable approximation.

Future work will extend the electron submodel from a 1D-MFAM to a two-dimensional magnetic-field-aligned mesh (2D-MFAM). This extension will enable the model to resolve electron particle and energy transport parallel to the magnetic field and to account for departures from complete thermalization along individual magnetic field lines.

## CRediT authorship contribution statement

**Xingdong Che:** Writing – review & editing, Writing – original draft, Visualization, Validation, Software, Methodology, Investigation, Formal analysis, Conceptualization; **Hong Li:** Writing – review & editing, Supervision, Software, Project administration, Methodology, Funding acquisition, Formal analysis; **Xin Guo:** Writing – review & editing, Visualization, Validation, Methodology, Investigation, Formal analysis; **Zhaoyu Wang:** Methodology, Investigation; **Xifeng Cao:** Methodology, Investigation; **Daren Yu:** Supervision, Funding acquisition.

## Data availability

Data will be made available on request.

## Declaration of competing interest

None.

## Acknowledgements

The work was supported by the National Natural Science Foundation of China under grant number U25B20235.